\documentclass[twocolumn,tighten]{aastex701}
\usepackage{amsmath}
\usepackage{tabularx} 
\usepackage{needspace}
\usepackage{appendix}
\usepackage{booktabs}     
\usepackage{graphicx}
\usepackage{hyperref}
\usepackage[table]{xcolor}
\usepackage{placeins}
\usepackage{enumitem}
\usepackage[normalem]{ulem}
\setlist{itemsep=2pt, parsep=0pt}

\newcommand\Msun{$\mathrm{M_{\odot}}$}

\definecolor{revisedmaroon}{RGB}{128, 0, 32} 
\newcommand{\edit}[1]{#1}

\newcommand{\age}{2.5\,Gyr} 
\newcommand{\totcat}{87} 
\newcommand{\totprimary}{71} 
\newcommand{\msto}{1.5} 
\newcommand{\totlcs}{81,498} 
\newcommand{\totstars}{11,055} 
\newcommand{\totnewvar}{\edit{26}} 
\newcommand{\totnewrot}{14} 
\newcommand{\bestppma}{33\,ppm} 
\newcommand{\bestppmb}{79\,ppm} 
\newcommand{\bestppmc}
{0.9\,millimag} 

\shortauthors{Soares-Furtado et al.}
\begin{document}

\title{Kepler Image-Subtracted Light Curves and Variable Star Catalog of NGC\,6819}

\author[orcid=0000-0001-7493-7419,sname='Soares-Furtado']{Melinda Soares-Furtado}
\affiliation{Department of Astronomy,  University of Wisconsin--Madison, 475 N.~Charter St., Madison, WI 53706-1390, USA}
\affiliation{Department of Physics, 2320 Chamberlin Hall, University of Wisconsin--Madison, 1150 University Avenue Madison, WI 53706-1390}
\email[show]{mmsoares@wisc.edu}  

\author[orcid=0009-0003-6919-5221,sname='Kuenzi']{Rianna Kuenzi}
\affiliation{Department of Astronomy,  University of Wisconsin--Madison, 475 N.~Charter St., Madison, WI 53706-1390, USA}
\email{rkuenzi@wisc.edu}  

\author[orcid=0000-0001-5928-7155,sname='McClure']{Rachel Lee McClure}
\affiliation{Department of Physics \& Astronomy, Weber State University, Ogden, UT 84408, USA}
\affiliation{Department of Astronomy,  University of Wisconsin--Madison, 475 N.~Charter St., Madison, WI 53706-1390, USA}
\email{rachelmcclure@weber.edu}  

\author[orcid=0009-0007-1123-0038,sname='Marston']{Nicholas T.~Marston}
\affiliation{Department of Astronomy,  University of Wisconsin--Madison, 475 N.~Charter St., Madison, WI 53706-1390, USA}
\email{nmarston@wisc.edu}

\author[orcid=0009-0006-7474-7889,sname='Linck']{Evan Linck}
\affiliation{Department of Astronomy,  University of Wisconsin--Madison, 475 N.~Charter St., Madison, WI 53706-1390, USA}
\email{elinck@wisc.edu}

\author[orcid=0000-0002-7130-2757,sname='Mathieu']{Robert D.~Mathieu}
\affiliation{Department of Astronomy,  University of Wisconsin--Madison, 475 N.~Charter St., Madison, WI 53706-1390, USA}
\email{mathieu@astro.wisc.edu}

\author[orcid=0000-0001-7246-5438,sname='Vanderburg']{Andrew Vanderburg}
\altaffiliation{Alfred P.~Sloan Research Fellow}
\affiliation{Center for Astrophysics | Harvard and Smithsonian, 60 Garden Street, Cambridge, MA 02138, USA}
\affiliation{Department of Physics \& Kavli Institute for Astrophysics and Space Research, Massachusetts Institute of Technology, Cambridge, MA 02139, USA}
\email{andrewv@mit.edu}

\author[orcid=0000-0001-8732-6166,sname='Hartman']{Joel D.~Hartman}
\affiliation{Department of Astrophysical Sciences, Princeton University, Princeton, NJ 08544, USA}
\email{jhartman@astro.princeton.edu}

\author{Kyle Cudworth}
\affiliation{Yerkes Observatory, Williams Bay, WI 53191, USA}
\email{kcudwort@uchicago.edu}

\author[orcid=0000-0002-5665-1879]{Robert Gagliano}
\affiliation{Amateur Astronomer, Glendale, Arizona, USA}
\email{astrowebdoc@gmail.com}

\author[orcid=0000-0003-3988-3245]{Thomas Jacobs}
\affiliation{Amateur Astronomer, Missouri City, Texas 77459 USA}
\email{tomjacobs128@gmail.com}

\author[orcid=0000-0002-2607-138X]{Martti H.~Kristiansen}
\affiliation{Brorfelde Observatory, Observator Gyldenkernes Vej 7, DK-4340 T\o{}ll\o{}se, Denmark}
\email{martti@outinto.space}

\author{Mark Omohundro}
\affiliation{Citizen Scientist, c/o Zooniverse, Department of Physics, University of Oxford, Denys Wilkinson Building, Keble Road, Oxford, OX1 3RH, UK}
\email{fly.fishing.heaven.pa@gmail.com}

\author[orcid=0000-0002-1637-2189]{Hans Martin Schwengeler}
\affiliation{Citizen Scientist, c/o Zooniverse, Bottmingen, Switzerland}
\email{hans.schwengeler@intergga.ch}

\begin{abstract}
Variable stars in open clusters are valuable probes of stellar evolution. They provide precise measurements of stellar properties, constrain cluster ages and distances, and trace the angular momentum evolution of stellar populations.
To advance these studies, we applied image subtraction and systematic reduction techniques to the NGC\,6819 \textit{Kepler} superstamp time-series data (Quarters 1-16), using the \textit{Gaia} DR3 catalog to identify the positions of individual sources. 
We produce \totlcs{} high-precision light curves for \totstars{} sources in the crowded field harboring the \age{} open cluster NGC\,6819. 
Our detrended light curves achieve a best root-mean-square precision of \bestppma{} (6.5-hour bins) for stars with \textit{Kepler} magnitudes $12-12.5$\,mag, falling to \bestppmb{} at $14-14.5$\,mag. 
Using \textit{Gaia} DR3 proper motions, and parallaxes, we distinguished likely cluster members from field stars. 
We identified and classified \totcat{} periodic variables that are potential members of NGC\,6819, including \totnewvar{} newly-discovered variables.
We make our light curves and variable classifications publicly available to enable further studies of stellar variability and angular momentum evolution in this intermediate-aged open cluster.
\end{abstract}

\msto

\section{Introduction}\label{sec:intro}

Stellar variability is a probe of fundamental astrophysics. Rotation periods measured from starspot modulation reflect stellar angular momentum evolution \citep[e.g.,][]{Barnes2007}. Asteroseismic oscillations yield precise measurements of masses, radii, and ages \citep[e.g.,][]{Chaplin2013}. Eclipsing binaries provide model-independent benchmarks for stellar evolution theory \citep[e.g.,][]{Torres2010}. 
To interpret these observables in an evolutionary context, they must be anchored to stellar populations with independently established ages and compositions. 

Open clusters provide exactly such reference sets: coeval stellar populations with common distances and metallicities that enable direct calibration of variability-based age-dating methods for field stars.
Gyrochronology is particularly dependent on cluster benchmarks to map the spin-down sequence over time, yet the current calibrators are unevenly distributed in age. 
Young clusters like the Pleiades ($\sim$125\,Myr) and Praesepe ($\sim$700\,Myr) have been extensively characterized, as has the older benchmark M67 ($\sim$4\,Gyr), however, intermediate-age clusters remain comparatively underexplored. 
This presents unique challenges. 
For example, the 1--3\,Gyr regime spans the transition from rapid early spin-down to the slower evolution that dominates at late times \citep[e.g.,][]{Barnes2007}.

NGC\,6819 is a particularly valuable benchmark cluster. At $\sim$2.5\,Gyr with near-solar metallicity \citep{Deliyannis2019}, it samples an age regime critical for gyrochronology and enables direct comparison with solar-like field stars. 
The cluster lies within the original \textit{Kepler} field \citep{Borucki2010}, providing four years of continuous, high-precision photometry for its $\sim$2,500 members (spanning the main sequence through the giant branch). Its moderate distance further enables spectroscopic follow-up for radial velocities, abundances, and binarity assessments. Notably, NGC\,6819 hosts several blue stragglers \citep{Mathieu2025}---objects that appear anomalously young and massive, likely products of mass transfer, mergers, or collisions. They are found within the classical instability strip, and their pulsations provide direct constraints on mass, radius, and internal structure. 

We present a comprehensive variability study of NGC\,6819 based on \textit{Kepler} Quarters 1--16. To mitigate stellar crowding, we employed a custom image subtraction pipeline and trend filtering algorithm. We extracted high-precision light curves for \totstars{} stars in the cluster field, including foreground and background sources. We determined cluster membership using \textit{Gaia} proper motions and radial velocities, then visually classified periodic variables among confirmed members. This analysis yields \totcat{} cluster variable stars, including newly identified variables and revised classifications for previously cataloged sources. In addition to gyrochronology, the resulting catalog supports studies of binary demographics, asteroseismology, and the evolutionary histories of blue stragglers. Our light curves are publicly available on the Mikulski Archive for Space Telescopes (MAST; \citealt{MAST}) and the full variable star catalog can be found on VizieR \citep{Vizier}. 

This paper is organized as follows. Section~\ref{sec:NGC6819} provides background information on the open cluster, including its key properties and relevance to variability studies. Section~\ref{sec:data} describes the data products used in our analysis. 
Section~\ref{sec:methods} describes our methods, including our processes for image subtraction and detrending, cluster membership, periodogram analyses, variable classification, and deblending.  
Section~\ref{sec:results} presents our findings, including an overview of the precision of our light curves and the resulting cluster variable star catalog.
Section~\ref{sec:disc} presents an analysis of our work in the context of previous studies.
In Section~\ref{sec:summary}, we summarize our key findings and conclusions.

\section{Overview of NGC 6819}\label{sec:NGC6819}

NGC\,6819 is a well-studied open cluster located in the Cygnus constellation at a distance of approximately 2.3--2.6\,kpc \citep{Ak2016,Cantat2018}. 
With a population of over 2,500 stellar members \citep{Zwicker_2024}, the system provides a rich laboratory for testing stellar evolution theories.

The cluster is estimated to be 2.3-2.5\,Gyr old \citep{Kalirai_2001,Meibom2015,Brewer2016}.
In our discussion of this system throughout the text, we adopt the 2.5\,Gyr age estimate. 
We estimate a main-sequence turn-off (MSTO) mass of \msto{}\,\Msun{}, corresponding to a \textit{Kepler} magnitude of approximately 14.5\,mag. 
The cluster has been subject to extensive ground-based observations \citep[e.g.,][]{mathieuWIYNOpenCluster2000,Yang_2012,Anthony_Twarog_2014,Milliman_2014, Deliyannis2019}.
High-resolution spectroscopic investigations indicate a near-solar metallicity, with \citet{Ak2016} reporting [Fe/H] values of $0.05\pm0.02$\,dex, which is in good agreement with prior studies by \citet{Milliman_2015}. 

Kinematically, NGC\,6819 orbits the thick Galactic disk with a period of 142\,Myr and an eccentricity of $e=0.06$ \citep{Ak2016}. Structural studies based on galactocentric $UBV$ observations resulted in a measured angular radius of $r=9'$, and a core radius of $r_c=3.65\pm0.38'$ \citep{Ak2016}. The core exhibits a stellar density of $13.18\pm0.46~\mathrm{stars\,arcmin}^{-2}$, while the surrounding background density is $5.98\pm0.45~\mathrm{stars\,arcmin}^{-2}$. 
A tidal radius of $40'$ has been reported \citep{Cantat2018}. 

\subsection{NGC 6819 Variability History}

The first variable investigation of NGC\,6819 was conducted by \citet{1967PZ.....16..191B}. While they identified potential candidates, their results were inconclusive. One initial candidate was later confirmed as a variable star by \citet{1971IBVS..606....1L}.
They reported evidence of a late-type irregular variable star situated 4\arcmin{} from the cluster center, classifying it as an M4-type giant based on objective prism plates and photoelectric measurements.
Subsequent studies, including those by \citet{1988AJ.....95..785K} and \citet{1991A&A...251...49M}, added a few more variable star candidates, including blue straggler stars exhibiting possible variability. However, these results were uncertain due to observational limitations.

A search for transiting exoplanets over a baseline of 19 nights significantly expanded the catalog of known variable stars, resulting in 25 new variable stars and 13 candidates \citep{2002MNRAS.330..737S}. Identified variables included eclipsing binaries, BY Draconis systems (rotational variability due to the presence of star spots), and a potential Cepheid variable (luminous, pulsating post-main-sequence star). An additional 141 variables---including 53 eclipsing binaries and 70 BY Draconis systems---were reported as unlikely members due to their distance from the center of the cluster \citep{Street2005}. Transiting exoplanets were not found in this or any other investigation to date, likely a consequence of limitations to sensitivity and blending with the cluster field's eclipsing binaries.  

Data from NASA's  \textit{Kepler Space Telescope}, in combination with the development of specialized data reduction pipelines, has enabled more detailed studies of variability in the cluster. The mission and corresponding data products are described in more detail in Section~\ref{subsec:Kepler}.
Kepler long-cadence light curves for stars in NGC\,6819 have been analyzed using the Increased Resolution Image Subtraction (IRIS) pipeline,\footnote{\url{https://archive.stsci.edu/hlsp/iris}} and compiled in the Kepler \texttt{IRIS} catalog. The Kepler \texttt{IRIS} catalog includes 3,808 stars in the NGC\,6819 superstamp (both members and nonmembers), and identified 254 targets as exhibiting variability without further classification \citep{Colman2022}.
More recently, \citet{Sanjayan2022} and collaborators used \textit{Kepler} data to identify 128 cluster variable stars, including 17 binary systems, 24 pulsational variables, and 82 rotational variables. In Section~\ref{sec:disc}, we describe how our variable catalog compares with these prior \textit{Kepler} variability analyses. 

\section{Data and Observations}\label{sec:data}  
\subsection{Kepler Observations of NGC 6819}\label{subsec:Kepler}  
NASA's \textit{Kepler Space Telescope} conducted a four-year primary mission (May 2009--May 2013) to identify transiting exoplanets in photometric time-series data \citep{Borucki2010,Brown2011}. 
The survey continuously observed a 115-square-degree field in the Cygnus-Lyra region, targeting approximately 150,000 stars \citep{Jenkins2010}. 
Each cadence consists of 6.02-second exposures with 0.52 seconds of readout time. 
The vast majority of targets were observed in long cadence mode, produced by combining 270 individual frames, for a total integration time of approximately 29 minutes.

Initially, the telescope achieved a precise pointing stability of 0.009\arcsec{} using four reaction wheels. However, two of these reaction wheels failed to reach their expected lifetimes. In May 2013, the loss of a second reaction wheel ended the primary \textit{Kepler} mission, ushering in the \textit{K2} extended mission, during which the spacecraft employed a novel two-wheel control scheme that leveraged solar radiation pressure for stabilization. For a more in-depth discussion of the \textit{Kepler} mission, we refer interested readers to \citet{Haas2010} and \cite{keplerhandbook}.

In the densely populated region containing the open cluster NGC\,6819, individual pixel stamps were grouped into larger \textit{superstamps}.  
The $200 \times 200$ pixel NGC\,6819 superstamp provides contiguous long-cadence monitoring.
The baseline of these observations are divided into 18 observing quarters (0-17), providing consistent time coverage for stellar monitoring.

To ensure consistent instrument performance across all light curves, we omit data from Quarter~0, the initial commissioning phase, which was obtained in coarse pointing mode and spanned a reduced baseline of only ${\sim}$10 days.
Detailed information about individual quarters is provided in \cite{keplerhandbook}.
Further, we omit data from Quarter~17 given the truncated baseline (due to the failure of a second reaction wheel) and the presence of gaps in the data due a safe-Mode event.\footnote{\url{https://archive.stsci.edu/kepler/release_notes/release_notes23/DataRelease_23_20131218.pdf}} 

\subsection{NGC\,6819 \& the WIYN Open Cluster Study}
The WIYN Open Cluster Study \citep[WOCS,][]{mathieuWIYNOpenCluster2000} produced 16,188 observations of 3,091 stars in the NGC\,6819 field (within a 30' radius) from 1998 June 12 through 2023 October 02 as part of a radial-velocity (RV) survey of rich open clusters. RV measurements and binary orbital solutions made prior to 2014 are published in \citet{tabethaholeWIYNOPENCLUSTER2009} and \citet{millimanWIYNOPENCLUSTER2014}. 
Spectra were taken using the Hydra Multi-Object Spectrograph on the WIYN 3.5-m telescope at Kitt Peak, Arizona at a resolution of approximately 20,000. 
Observations consisted of three 20- to 40-minute science exposures, a 200-second dome flat, and two 300-second thorium-argon emission lamp spectra in a 250\,\AA{} window centered on 5,125\,\AA{}. 
RVs were derived by cross-correlating each spectrum with an observed solar template spectrum using the \textsc{IRAF} cross-correlation function routine \textsc{fxcor}. 
WOCS RVs have a precision of $\mathrm{0.4\; km\;s^{-1}}$ for stars with projected rotational velocities ($v\sin{i}$) less than ($\mathrm{10\; km\; s^{-1}}$).
Precision decreases for stars with higher $v\sin{i}$, following: 
\begin{equation}
    \sigma = 0.38 + 0.012 [v \sin(i)] \; \mathrm{km} \, \mathrm{s}^{-1}
\end{equation}
\citep{gellerWIYNOPENCLUSTER2010a}.
Full details of the telescope, instrument, observing procedure, reduction process, and stellar sample are listed in the following publications:  \citet{gellerWIYNOPENCLUSTER2008, tabethaholeWIYNOPENCLUSTER2009, gellerWIYNOPENCLUSTER2010a}. 

\section{Methods}\label{sec:methods}

\subsection{Image Subtraction and Detrending Procedures}\label{subsec:imsub} 
In densely populated stellar fields, traditional aperture or PSF-fitting photometry techniques are often compromised by source blending and crowding. 
To address these challenges, we applied an image subtraction process following the algorithm described by \citet{Alard1998} and \cite{Alard2000}. 

Through the construction of a stable, high-quality reference image, which is subtracted from all individual cadence images, image subtraction isolates the time-varying flux contributions. This approach effectively removes the static background field of non-variable stars, reduces blending from neighboring sources, and improves the detectability and precision of subtle brightness variations.

Our software pipeline follows the methodology described in \citet{Huang2015} and \citet{SoaresFurtado2017}. The key steps are briefly summarized below:

\begin{enumerate}
    \item \textbf{Superstamp Assembly:}  
    For each cadence, we combine the Target Pixel Files (TPFs) from the relevant \textit{Kepler} module into a single superstamp. We employed only observations that were not flagged for data quality concerns, which helps to ensure reliable time-series measurements over multiple quarters.

    \item \textbf{Source Extraction and Astrometry:}  
    Using the \textit{Gaia} DR3 catalog, we identify stars and establish astrometric transformations between \textit{Gaia} and \textit{Kepler} coordinates. The high spatial resolution and wide brightness coverage of \textit{Gaia} DR3 ensure precise and stable astrometric references.

    \item \textbf{Designation of an Astrometric Reference Frame:}  
    A sharp superstamp image taken when the telescope’s pointing was closest to the median direction of the observing campaign is selected as the “astrometric reference frame.” This choice helps minimize systematic offsets when aligning all other frames to this reference.

    \item \textbf{Spatial Alignment:}  
    All cadences are spatially transformed into this common reference frame, minimizing spacecraft jitter and maintaining consistency for subsequent analysis.

    \item \textbf{Master Photometric Frame:}  
    A “master photometric reference frame” is constructed by median-stacking all spatially aligned \textit{Kepler} frames.

    \item \textbf{Difference Imaging:}  
    Each cadence is subtracted from this master frame, isolating variable signals by removing the static background of non-varying stars.

    \item \textbf{Photometry:}  
    Time-dependent flux variations are measured directly from the difference images.

    \item \textbf{Light Curve Assembly:}  
    Light curves are constructed for all identified sources.

    \item \textbf{Systematic Noise Removal:}  
    The Trend Filtering Algorithm (TFA; \citealt{Kovacs2005}) is applied to remove residual systematics and enhance light-curve precision.
\end{enumerate}

Following this approach, we generated \totlcs{} image-subtracted light curves from  \totstars{}  distinct stars, with each light curve corresponding to data taken during a single \textit{Kepler} quarter. 
The resulting image-subtracted light curves are analogous to those generated in \citet{SoaresFurtado2017}, which focused on a variable analysis of the for the M35 and NGC~2158 open clusters.
These light curves have been made publicly available on the Barbara A.~Mikulski Archive for Space Telescopes (MAST) data archive.

\subsection{NGC 6819 Membership Analysis}\label{sec:membership}

We identified likely NGC\,6819 cluster members using data from the \textit{Gaia} eDR3 survey \citep{GaiaCollaboration2021}. To begin this process, we identified all \textit{Gaia} targets within a 23\arcmin\ radius of the cluster center (RA=295.327$^{\circ}$, DEC=40.190$^{\circ}$; \citealt{CantatGaudin2020}), where the cluster's proper-motion distribution remains distinct from that of the surrounding field stars. This initial selection yielded 44,342 sources.  

To further refine our sample, we applied geometric distance estimates from \citet{BailerJones2021}, retaining only sources that overlapped with the cluster’s distance of 2.3\,kpc (within their respective $\pm3\sigma$ intervals; \citealt{Sanjayan2022}).
This step removed an additional 10\% of the sample.

For the remaining sources, we modeled the proper-motion distribution as a mixture of two two-dimensional Gaussians: one representing the cluster members and the other the field population. 
Recognizing that proper-motion errors are only reliably small ($<0.1$\,mas) for stars with \textit{Kepler} magnitudes brighter than approximately $17$\,mag, we divided the sample into several magnitude bins with consistent error properties.
Within each bin, we fit the Gaussian models to the proper-motion data and calculated membership probabilities for each target based on the relative likelihood of belonging to the cluster versus the field.
Each star’s membership probability was then calculated as the ratio of the cluster Gaussian’s density at that star’s proper motion to the sum of the cluster and field Gaussian densities.
For stars with \textit{Kepler} magnitudes fainter than 19.6\,mag, the cluster's proper-motion distribution becomes indistinguishable from the field, preventing reliable membership classification. These sources were assigned a \texttt{NaN} probability.

\begin{figure}[htb!]
    \centering    
    \plotone{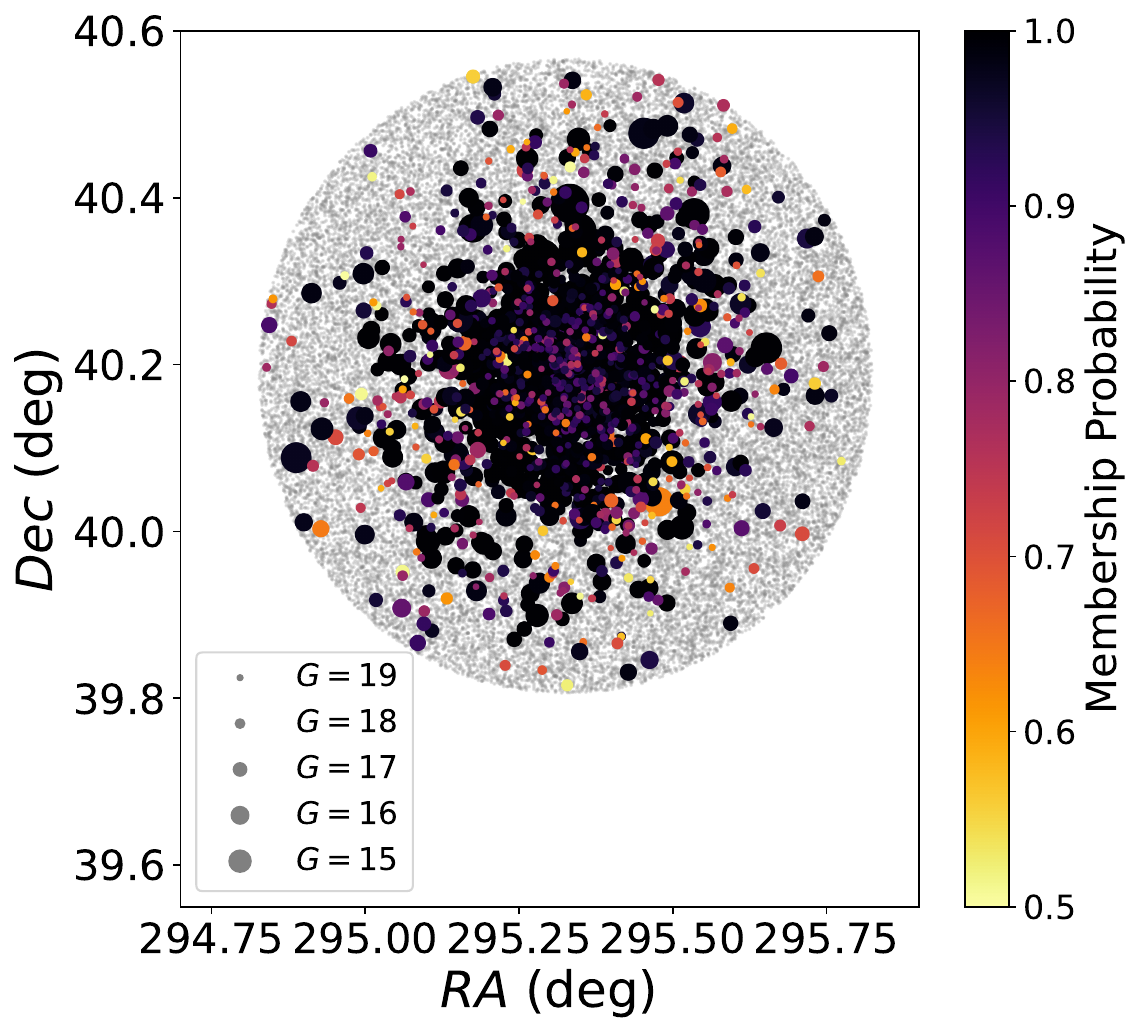}
    \plotone{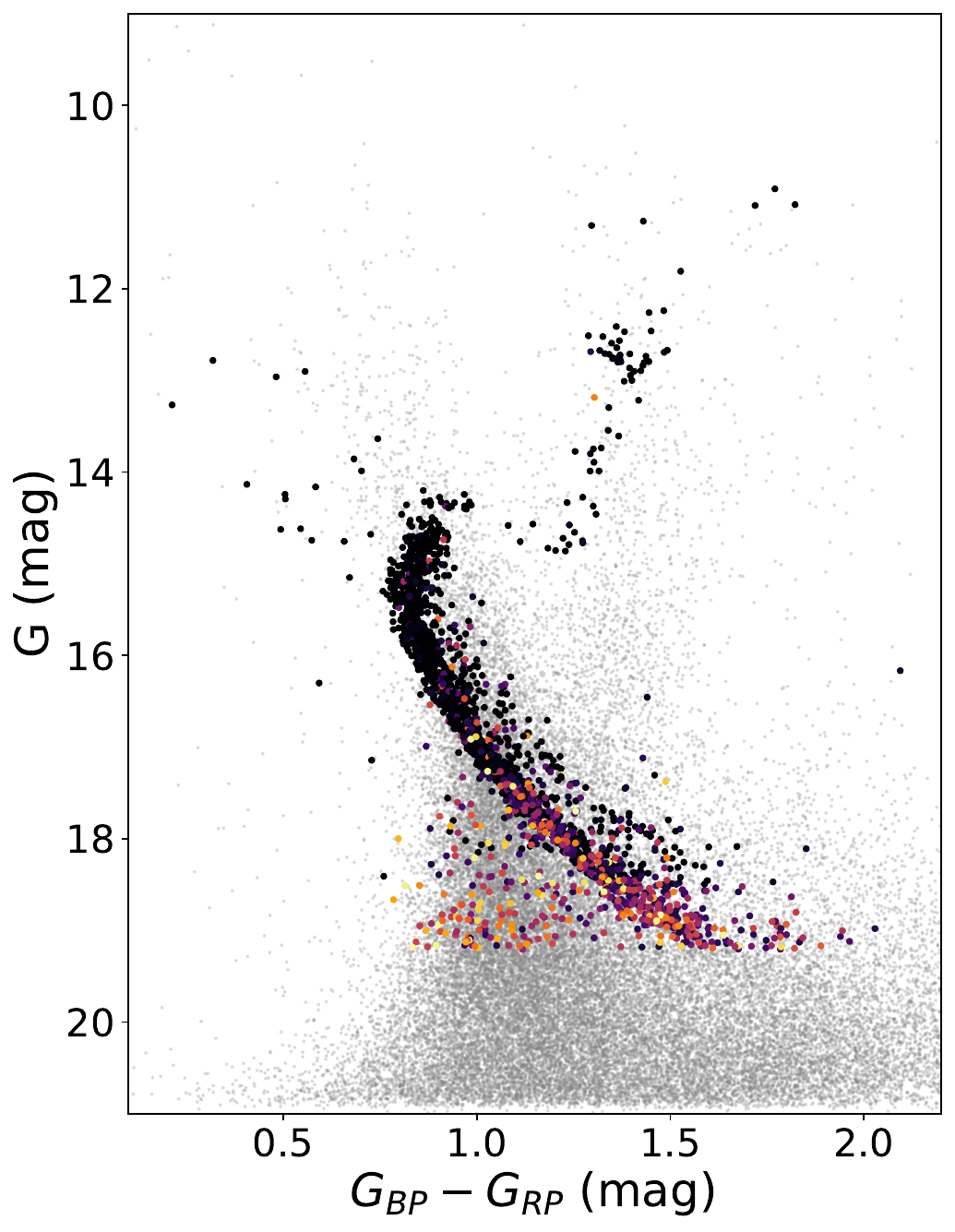}
    \caption{\textbf{Top:} Spatial distribution of likely NGC\,6819 members ($P \geq 0.5$) identified from \textit{Gaia} eDR3 data within 23\arcmin\ of the cluster center. Color indicates cluster membership probability. Marker size scales with brightness. Greyscale points illustrate all \textit{Gaia} sources in the field for comparison.
    \textbf{Bottom:} Color-magnitude diagram of likely NGC\,6819 members ($P \geq 0.5$) compared to all \textit{Gaia} sources within a 23\arcmin{} radius. For reference, the bulk reddening correction for the cluster is $E(\mathrm{BP}-\mathrm{RP}) = 0.18$\,mag.}
\label{fig:combined}
\end{figure}

We adopt a membership probability threshold of 50\%, which provides a balance between completeness and contamination while maintaining consistency with prior NGC~6819 studies \citep[e.g.,][]{tabethaholeWIYNOPENCLUSTER2009, Platais2013,Sanjayan2022}.
We identified 2,354 likely cluster members within the 23\arcmin{} radius. The spatial distribution of likely members ($P \geq 0.5$) is shown in the top panel of Figure~\ref{fig:combined}, revealing a concentration toward the cluster core.
Approximately half of the NGC\,6819 likely members are located within 7.7\arcmin{} (5.33\,pc) of the cluster center, confirming that our selection captures both the core and halo members of NGC\,6819.

In the bottom panel of Figure~\ref{fig:combined}, we illustrate the location of likely NGC\,6819 members on a color-magnitude diagram (CMD).
These targets make up a well-defined main sequence, a prominent binary sequence, a distinct giant branch with a red clump, and a population of blue straggler stars. 
Comparing our membership analysis with results from \citet{CantatGaudin2020}, we found over 90\% agreement for stars with \textit{Kepler} magnitudes brighter than 18\,mag at both the 50\% and 90\% membership probability thresholds. For fainter sources, completeness decreased due to crowding effects \citep[][]{Fabricius2021}. At the outer edges of our 23\arcmin{} selection radius, contamination from field stars increased but remained well-controlled by our chosen parallax and proper-motion analysis.

\subsection{Periodic Variability Analyses}\label{subsec:periodograms}
We identified periodic variability in our image-subtracted, trend-filtered light curves from an analysis incorporating the following three periodogram algorithms:

\begin{enumerate} 
\item \textbf{Box-fitting Least Squares (BLS):} \citep{Kovacs2002}---optimized for detecting box-shaped dips characteristic of eclipsing binaries (EBs) and transiting exoplanets. 
\item \textbf{Generalized Lomb-Scargle (LS):} \citep{Scargle1982, Zechmeister2009}---particularly sensitive to sinusoidal signals, commonly associated with stellar rotation or pulsations. 
\item \textbf{Phase Dispersion Minimization (PDM):} \citep{Stellingwerf1978}---well-suited for identifying non-sinusoidal periodicities or more complex light curve morphologies. 
\end{enumerate}

Quarters were investigated independently rather than stitching time-series observations together across the overall mission baseline.
Given the ${\approx}$93-day roll of the spacecraft, in each quarter a given target will fall on a different CCD module with its own pixel response function, focus, and noise characteristics. This results in quarter-specific flux offsets and systematics \citep{keplerhandbook}.
We set the maximum period for each target in a given quarter to half of that quarter’s temporal baseline.  This guarantees that any true photometric signal will complete at least two full cycles within a single observation quarter.
We sampled all periodograms at the Rayleigh frequency resolution 
($\Delta f = 1/T$, where $T$ is the total duration) up to the Nyquist 
frequency limit set by the observing cadence.

While our variable catalog intended to catalog only likely members of NGC\,6819, it was necessary to also investigate neighboring field stars for possible blending. 
Using the \texttt{astrobase} Python package \citep{Bhatti2018}, we analyzed the light curves of likely NGC\,6819 members and all nearby neighbors identified within a 30-pixel (2\,arcminute) radius. This search radius encompasses the majority of potential contaminating sources and enables verification that targets are the true variability source, as signals diminish with distance from the target star.
Three team members visually inspected each source, flagging ambiguous cases for further review. This included sources with low SNR, uncertain or non-repeating periodicities, or very low photometric amplitudes. 
We documented unresolved ambiguities in our catalog comments for transparency.

To decide which power spectrum peaks were robust, we required at least one of the following criteria to be true for at least a single quarter of data: Lomb Scargle False Alarm Probability (FAP) $<0.01$ or a Box-Least‐Squares Signal‐Detection‐Efficiency (SDE) $>6$. 
We then visually inspected the three strongest signals from each periodogram analysis by phase-folding the light curves and evaluating their morphology and amplitude, taking into account the known stellar parameters.
For instance, we considered whether the CMD position of the star was within the instability strip and if its power spectrum displayed comb-like frequency spacing indicative of solar-like oscillators.
We also cross-matched our targets with the WOCS database, investigating the RV scatter and orbital solutions (when available). Taking this information into account, we identified the most reliable period and refined our variable classification. 

To assign uncertainties to the adopted periods, we computed three independent error estimates for each target and adopted the largest of the three. 
First, since periods were measured independently in each quarter, we calculated the median absolute deviation (MAD) of the periods across all quarters, capturing the empirical scatter arising from evolving spot patterns, quarter-specific systematics, and differences in noise levels.
Second, we imposed a baseline-resolution error floor corresponding to the
Rayleigh frequency limit.  
Propagating this into period space yields a minimum
uncertainty of $\Delta P_{\mathrm{Rayleigh}} = P^{2}\Delta f$.  
Third, we included a quantized cadence error floor to prevent
sub-cadence precision claims. We adopted $\Delta P_{\mathrm{cadence}} = \Delta t/\sqrt{12}$, where $\Delta t$ is the
median light curve sampling interval.  
Our reported period uncertainty for each star is given by:
\begin{equation}
    \sigma_{P} = \max\left(\mathrm{MAD},\, \Delta P_{\mathrm{Rayleigh}},\,
    \Delta P_{\mathrm{cadence}}\right),
\end{equation}
ensuring that our reported uncertainties account for empirical and
theoretical resolution limits.

\subsection{By-Eye Variable Flagging by the Visual Survey Group}

In an effort to optimize the identification of variable stars, our light curves were also shared with the \textit{Visual Survey Group} (VSG). Several VSG collaborators conducted a systematic visual inspection to flag candidate periodic variable stars.
The VSG is responsible for variable discoveries using \textit{Kepler/K2} and the \textit{TESS} \citep[e.g.,][]{Schmitt2016,Mann2020,Powell2021,Kristiansen2022,Capistrant2024}. The VSG team used a Windows-based publicly available software system known as \texttt{LcTools} \citep{schmitt2019lctools,schmitt2021lctools}. \texttt{LcTools} provides users with three principal tools: \texttt{LcGenerator} to create normalized light curves in bulk, \texttt{LcSignalFinder} to automatically find and record candidate signals (periodic and aperiodic) in large sets of light curves, and \texttt{LcViewer} to visually inspect light curves for candidate signals.

\subsection{Classification of Periodic Variables}
\label{subsec:var_class}

Once we identified reliable periods, we separated our population of periodic variables into two categories:  
\begin{itemize}
    \item[] Class A: Light curves exhibiting clear \edit{eclipse-like dimming}---indicative of eclipsing binaries or potential transiting exoplanets. The light curve morphology can exhibit v-shaped or box-shaped dimming events.  
    \item[] Class B: Light curves exhibiting \edit{sinusoidal or complex morphology}---suggestive of rotating spotted stars, pulsational variables, or other non-transiting variability.
\end{itemize}

For box-like dimming signals, we assessed whether the transit depth, duration, and orbital separation were consistent with a star-planet system. To reduce false positives, we incorporated spectroscopic data, \textit{Gaia} parallaxes, stellar densities, and relevant cluster parameters (age, metallicity). 

To distinguish EBs from transiting exoplanets, we looked for a secondary eclipse or other morphological hallmarks of binary systems. 
Notably, some close-orbiting EBs exhibit highly sinusoidal light curves and were initially placed in Class B based on their photometric morphology; subsequent investigation of stellar parameters, RV variability, and other diagnostics ensured a more accurate final classification.

Whenever available, we cross-checked the photometric variability with published and new orbital solutions derived from WOCS RV data. To calculate orbital solutions, we relied on the open-source software \texttt{thejoker} \citep{2017ApJ...837...20P}. When these RV data were not available, we relied on the shape of the light curve by eye.

Class B periodic variables were further classified using their periods, light curve shapes, variability amplitudes (from Fourier fits), and stellar parameters (temperature, radius, luminosity). For rotating variables, we ensured that their rotational velocity did not exceed the break-up speed and that the star's effective temperatures were below 6250\,K---above which star spot activity is uncommon in dwarfs. 

Our variable classification designations are based on those outlined by the American Association of Variable Star Observers (AAVSO) International Variable Star Index \citep[VSX;][]{Watson2006,Watson2017} and \citet{Karttunen2017}. 
Within our period and amplitude bounds, we considered a wide range of pulsational variables: $\delta$ Scuti, RR Lyrae, $\gamma$ Doradus, Cepheids, $\beta$ Cephei, and slowly pulsating B-type stars (SPB). Classes outside our investigated period range were excluded. 

Our classification pipeline incorporated a hierarchical decision tree that sequentially eliminated incompatible classes based on four primary diagnostics:
\begin{enumerate}
    \item \textbf{Light curve morphology:} We distinguish between symmetric sinusoidal variations (characteristic rotational variables and single-period pulsators), asymmetric sawtooth patterns with steep rising branches (RR Lyrae subtype ab, high-amplitude $\delta$ Scuti), and complex multi-periodic behavior ($\gamma$ Doradus, SPB stars exhibiting multiple g-modes).
        
    \item \textbf{Period distribution:} Ultra-short periods ($<$0.1 d) indicate $\delta$ Scuti or $\beta$ Cephei pulsators; intermediate periods (0.2--1.2 d) encompass RR Lyrae variables; periods of 0.3--3 d suggest $\gamma$ Doradus; and longer periods (2--30 d) indicate Cepheids. Spectral-class dependent rotational variable periods bounds were also imposed. 
    
    \item \textbf{Color-magnitude diagram position:} Stellar location on the CMD provides critical constraints---$\delta$ Scuti occupy the lower instability strip (A--F spectral types), RR Lyrae populate the horizontal branch, Cepheids trace the upper instability strip as evolved supergiants, while $\beta$ Cephei and SPB stars are confined to the hot main sequence (B-type stars). Rotational variables are predominantly found along the main sequence.
\end{enumerate}
Ambiguous cases where multiple classifications remain viable were assigned a ``misc" classification (numerical class 4 in the catalog).

\subsection{Deblending}\label{subsec:deblend}

To identify the true source of a variable signal in crowded regions, we compared each star’s already‑measured period(s) (see Section~\ref{subsec:periodograms}) with those of all neighbors within a 30‑pixel radius.  Any target whose period or its harmonics did not appear in a neighbor's periodogram was accepted as the primary variable, requiring no further deblending analysis.

For cluster variables in blended groups, additional steps were required to identify the primary source of variability. 
In the case of EB candidates, we once again leveraged the extensive RV data from the WOCS program. 
These RV measurements taken over a long temporal baseline included targets with well-defined orbital solutions, enabling our team to distinguish eclipsing binary primaries with improved spatial resolution ($3\arcsec$ fiber diameter). 
This was particularly valuable given the slightly larger pixel size of the \textit{Kepler} CCDs ($3.98\arcsec$/pixel), as well as instances of significant signal leakage (flux leakage into neighboring pixels) in crowded fields.
By incorporating WOCS RV measurements, we more reliably isolated the true source of variability within blended groups.
Our variable catalog distinguishes cases where the WOCS orbital solution corroborates the photometric variability signature of a given target. 

For variables exhibiting pulsational-like variability, we employed additional methods to discern the primary source within blended groups. Pulsating variables are typically found within the instability strip of the color-magnitude diagram. To identify the likely primary target, we examined the extinction- and reddening-corrected CMD positions of all sources within a blended group. 
The primary source of pulsational variability is expected to occupy a position consistent with the instability strip, while neighboring contaminants outside this region were excluded as likely secondary sources.

It is important to note that in many cases it was not possible to confidently identify the primary source of variability within a blended group. For such blended systems, we label the targets in our variable catalog as ``ambiguous blends" to indicate the unresolved nature of the variability source. These ambiguous cases highlight the limitations imposed by \textit{Kepler}'s spatial resolution and the challenges of analyzing crowded fields.

\section{Results}\label{sec:results}

\subsection{Light-Curve Photometric Precision} 
\label{sec:LCs}
\begin{figure*}[t!]            
    \includegraphics[width=\linewidth]{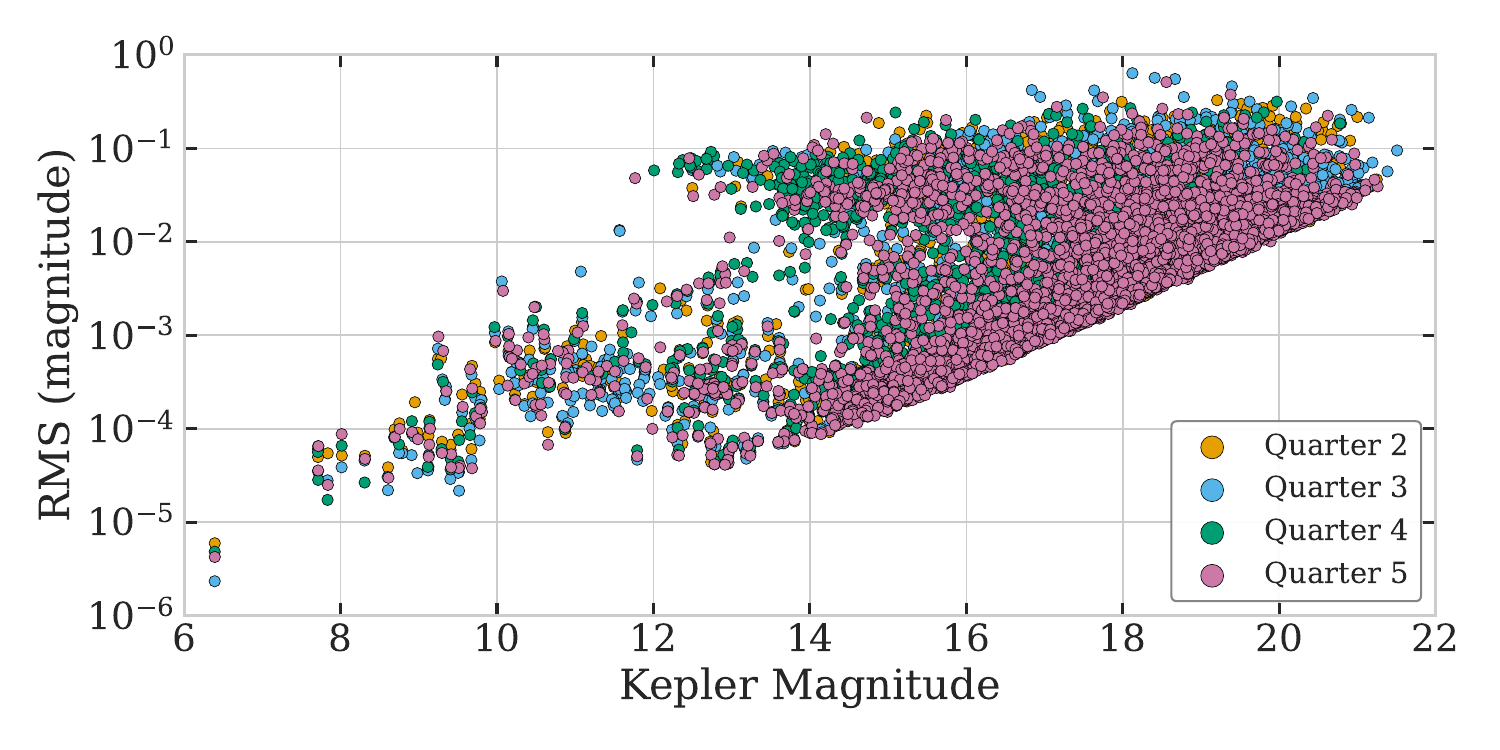}
    \caption{Photometric precision (RMS scatter) as a function of \textit{Kepler} magnitude for our detrended light curves. A subset of \edit{consecutive} quarters \edit{all falling on different channels} are shown. While these quarters are selected arbitrarily, they are reflective of the overall photometric precision achieved. This figure demonstrates the sub-millimagnitude precision ($\lesssim10^{-4}$\,mag) achieved for the bright stars and the increase in photometric noise for fainter stars. Many of the stars above the photometric noise floor are variable stars.}
    \label{fig:rmsscatter}
\end{figure*}

To quantify the precision of our detrended, TFA-corrected, image‐subtracted light curves, we computed the running‐window root‐mean‐square (RMS) over a 6.5-hr window (13 cadences) for every source in a given quarter.
Figure~\ref{fig:rmsscatter} illustrates the RMS scatter as a function of \textit{Kepler} magnitude, for TFA‐corrected light curves in several \edit{consecutive} quarters. 
As discussed in Section~\ref{subsec:periodograms}, and evident in this figure, quarter‐specific systematics cause the photometric precision of each source’s light curve to vary from one quarter to the next \citep{keplerhandbook}.
However, the quarters illustrated here are reflective of the spread in photometric precision across all quarters. 

For stars with \textit{Kepler} magnitudes of $12.0-12.5$\,mag, we achieve a best 6.5-hr RMS of \bestppma{}, rising to \bestppmb{} for $14.0-14.5$\,mag.  
Fainter stars are dominated by photon noise, with RMS values of \bestppmc{} for stars of $17.0-17.5$\,mag.
Faint sources $>15.5$\,mag with anomalously low RMS values (below the photon noise limit) are sources for which the catalog magnitude is brighter than the true magnitude on the photo-reference image, leading to underestimating the amplitude of photometric variation.

Our results achieve significantly improved photometric precision from the analogous \textit{K2} analysis by \citet{SoaresFurtado2017} due to the improved pointing stability of the original \textit{Kepler mission}. 
Specifically, stars in the $12.00-12.25$\,mag range exhibit a 4.4-fold improvement in photometric precision, while those in the $14.00-14.25$\,mag range show a 2.7-fold improvement.
Further, our results closely track the \textit{Kepler} superstamp image-subtracted light curve analyses from \citet{Colman2022}. 

\subsubsection{Comparison with the \texttt{IRIS} Light Curves}\label{sec:disc}
We conduct a direct comparison of photometric precision between the light curves produced in our study and the \texttt{IRIS} data products \citep{Colman2022}, which likewise use image subtraction to isolate variability in crowded fields. 
Both pipelines perform aperture photometry on the residual frames to extract stellar light curves. However, the pipelines diverge in regard to how they handle image alignment, PSF matching, and systematics removal.

We identified stars for which light-curve products exist in both samples. 
Since we used the \textit{Gaia} eDR3 catalog to identify source centroids for individual targets rather than the \textit{Kepler} Input Catalog (KIC) used by \citet{Colman2022} our parent sample is substantially larger. 
\textit{Gaia} eDR3 provides precise positions for nearly two billion sources, whereas the KIC is limited to approximately 13 million targets.
This larger source catalog allowed us to extract high-quality light curves for a significantly larger and more complete stellar population within NGC\,6819.
As a result, our dataset includes many sources that were inaccessible in previous variability studies, particularly in crowded regions where accurate source separation depends critically on precise astrometry.

Of the individual light curves we extracted for a given quarter, approximately 6\% were found to have matching \texttt{IRIS} light curves available on MAST.
For each match, we converted the \texttt{IRIS} fluxes to magnitudes.
We then applied a one-day high-pass filter to the light curves to remove long-term variability. This ensures that our subsequent noise estimates and period searches genuinely reflect the high-frequency noise floor and short-period signals.
We likewise sigma-clip our light curves at a $3\sigma$ threshold in three iterations to match the \texttt{IRIS} pipeline and ensure a more directly comparable analysis. 

We then computed the 6.5-hr rolling median RMS of each light curve to obtain a robust, directly comparable measure of photometric scatter. 
In the comparison of RMS scatter between the two detrending methods across all examined quarters and channels, we see that the two pipelines generally result in comparable photometric precision. 

\begin{figure}[tbh!]
\centering
\plotone{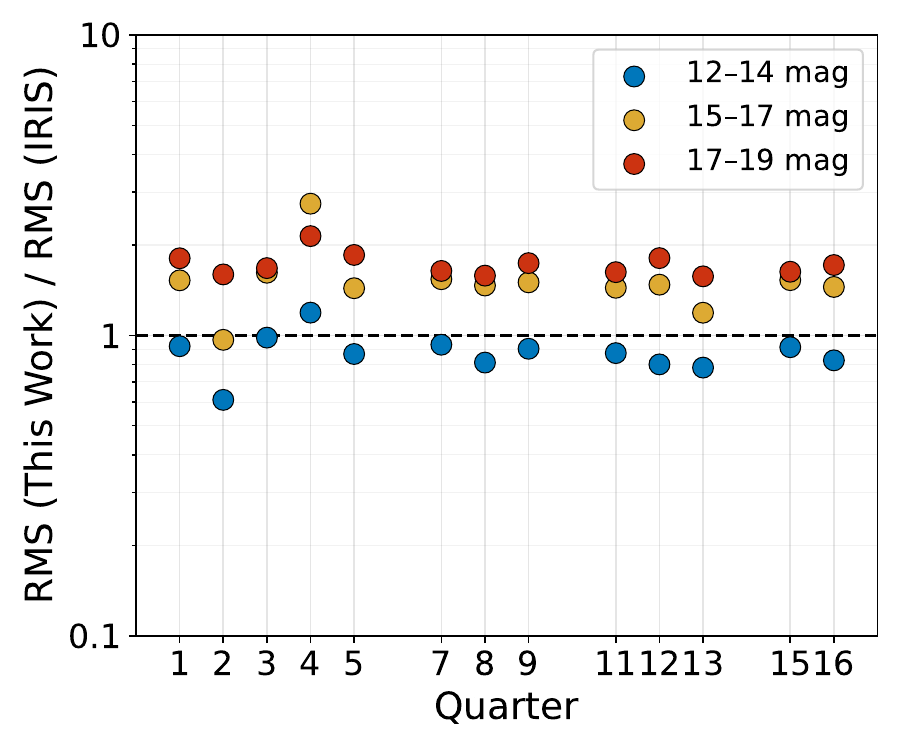}
\caption{Median value of the ratio of the RMS scatter of our light curves to that from the \texttt{IRIS} pipeline \citep{Colman2022} as a function of \textit{Kepler} quarter for targets in differing brightness bins. The RMS ratio is computed using a 6.5-hour (13-point) rolling median on the magnitudes, with a dashed line at unity indicating equal photometric precision.}
\label{fig:rms_scatter}
\end{figure}

In Figure~\ref{fig:rms_scatter}, we show the median ratio of RMS scatter between our light curves and the corresponding \texttt{IRIS} products as a function of \textit{Kepler} quarter, with results shown separately for intermediate-brightness stars ($12 \le K_{p} \le 14$\,mag) and for faint stars of $15 \le K_{p} \le 17$\,mag and $17 \le K_{p} \le 19$\,mag.
For stars of intermediate brightness, our reduction more consistently yields marginally improved photometric precision.
In the faint, photon‐noise-dominated regime, the \texttt{IRIS} pipeline more consistently yields improved photometric precision.

The difference in photometric precision performance likely stems from differences in PSF matching and removal of systematics under different signal-to-noise regimes. Our light curves are produced using convolution kernels for PSF matching with the application of TFA corrections to strip out common trends---an approach that
delivers exceptionally low scatter for bright stars but becomes problematic for fainter stars where the method overfits to noise patterns. 
By integrating the strengths of both of these high-level science products, one can achieve uniformly high photometric fidelity over the entire magnitude range in the NGC\,6819 field. 

\edit{This reduced faint-end precision has a limited and well-characterized impact on our catalog. 
Since we identify variables through periodogram significance thresholds in tandem with visual vetting, the dominant consequence is one of completeness rather than reliability. 
At $15-19$\,mag we are sensitive only to variables whose amplitudes exceed the local noise floor, therefore, the lowest-amplitude faint variables recoverable from the IRIS products may be absent from our sample. 
The faint variables we do recover are detected at high periodogram significance and have peak-to-peak amplitudes that typically exceed the local 6.5-hr RMS by an order of magnitude or more (Table~\ref{bigtable}).
Their classifications rest on CMD position,  light curve and power-spectrum structure, and, when available, WOCS RV solutions (Section~\ref{subsec:var_class}). 
We quantify the resulting completeness difference directly in Section~\ref{subsubsec:comparison}.}

\subsection{Variable Catalog Census} 
\label{sec:census}

\begin{figure*}[htb!]
    \centering
\includegraphics[width=0.97\textwidth]{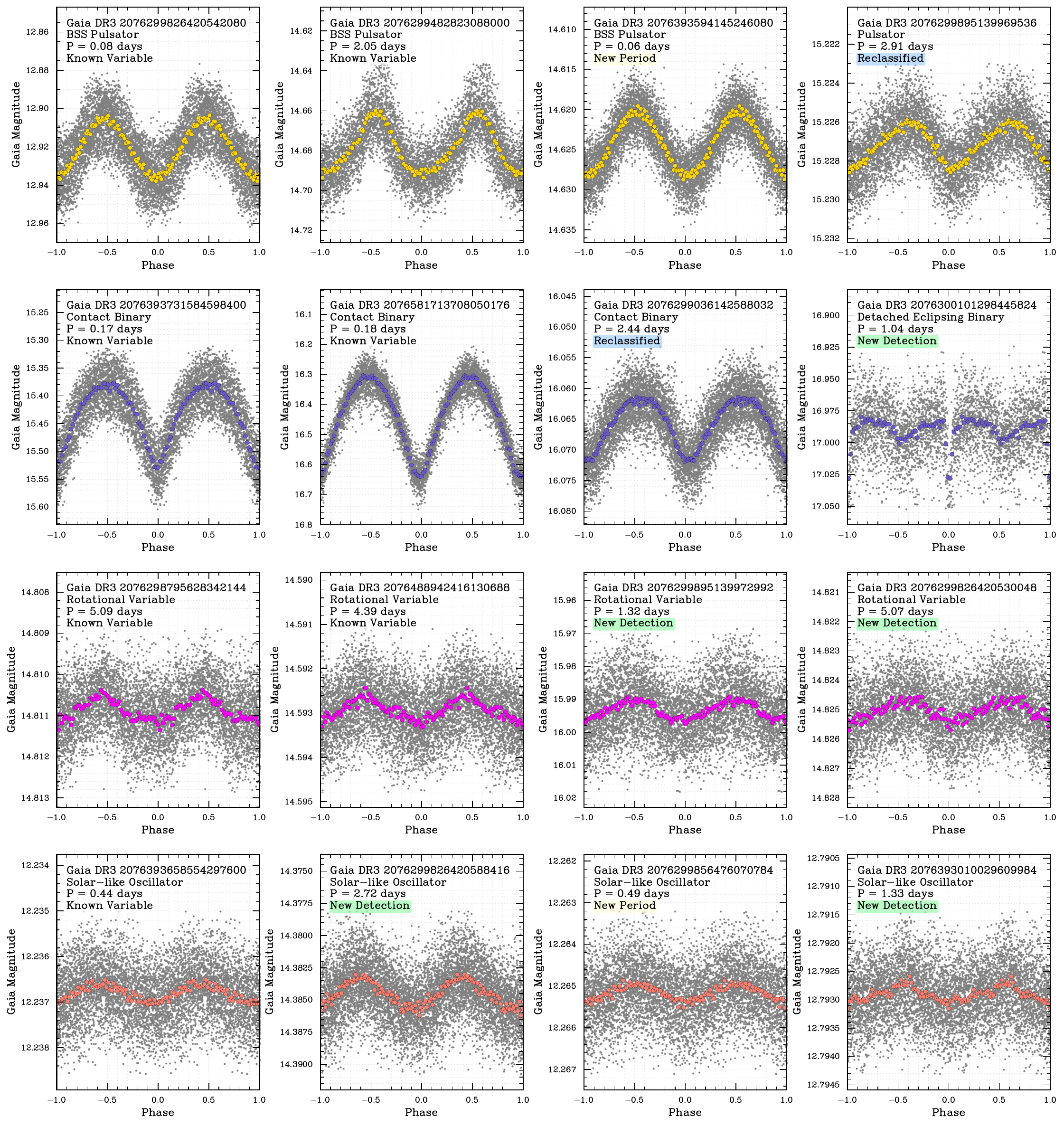}
    \caption{
    Phase-folded light curves for a sample of variable stars in our catalog, organized by variability class: pulsators (top row), binary-induced variability (second row), rotational variables (third row), and solar-like oscillators (bottom row). Each panel illustrates the best-fit phase-folded period from the \textit{Kepler} quarter with the strongest periodogram power. Binned data are color-coded by variability class: pulsational variables (yellow), eclipsing binaries (blue), rotational variables (magenta), and solar-like oscillators (salmon). As this figure makes clear, light-curve morphology alone is frequently insufficient to assign a variability class. F
    or this reason we do not classify on morphology alone, but combine it with color-magnitude diagram placement, the power spectrum structure, and radial-velocity orbital solutions and scatter. Labels indicate the status of each source: previously known variables with confirmed classifications, reclassified variables (highlighted in blue), and new detections (highlighted in green).}    
    \label{fig:example_lcs}
    \end{figure*}

We identify \totcat{} periodic variable sources in the cluster NGC\,6819 (membership probability $>0.5$).
In Figure~\ref{fig:example_lcs}, we present a selection of phase-folded light curves from our catalog, illustrating the range of variability types identified.
The figure is organized by variability class, with each row corresponding to a different type: pulsators (top), binary-driven variability (second row), rotational variables (third row), and solar-like oscillators (bottom). Each row includes both previously known variables, as well as new discoveries (when possible) or reclassifications from our analysis. This figure is intended to demonstrate the quality of our light curves across the full range of variability amplitudes and periods in the catalog.

The periodic variables identified in our analysis include the following:
\begin{itemize}
    \item 28 rotational variables
    \item 25 solar-like oscillators
    \item \edit{18 stars with variability driven by binarity}
    \item Seven sources with variability that could not be classified (miscellaneous variables)
    \item Five main-sequence pulsational variables, likely of $\gamma$ Doradus and $\delta$ Scuti types
    \item Four blue straggler stars exhibiting pulsational variability
    \item One previously identified detached eclipsing binary (WOCS\,24009) that was found by \citet{Brewer2016} to be a triple-lined spectroscopic binary. Using RVs and eclipse timing variations they found that the brightest star is physically orbiting the inner eclipsing binary.
\end{itemize}
\edit{A subset of columns from our variable catalog is presented in Table~\ref{bigtable} in the Appendix}.
While we report \textit{Gaia} $G$ magnitudes in our catalog tables, the photometric precision values are fundamentally determined by \textit{Kepler} observations, which depend on how bright each star appears to the \textit{Kepler} instrument. We adopt \textit{Gaia} magnitudes because they provide uniformly calibrated millimagnitude precision for all sources and align with our \textit{Gaia} eDR3-based membership analysis. Since \textit{Kepler} and \textit{Gaia} $G$ magnitudes typically differ by only $\sim$0.1\,mag for our targets, the brightness ranges are effectively equivalent when interpreting photometric performance.

\subsection{New Variable Star Discoveries}

Of the \totcat{} periodic variables identified, \totnewvar{} are new detections not previously reported in the literature. 
The new detections span multiple variability classes: 14 rotational variables, \edit{six solar-like oscillators, four variables of indistinguishable type, and two eclipsing binaries.}
We additionally provide new classifications for \edit{13} previously known variables and report an updated period \edit{for three sources.}
The reclassifications primarily affect pulsators (\edit{five} sources formerly classified as rotational variables or eclipsing binaries) and binary systems (seven sources). We distinguish between new detections (flag [1]), improved classifications of known variables (flag [2]), and refined period measurements (flag [3]) in our catalog. Variables with only flag [2] or [3] are not counted among the \totnewvar{} new discoveries since they were previously detected.

From our deblending analysis (Section~\ref{subsec:deblend}), we found that \totprimary{} of our \totcat{} variable stars were the clear primary source among all stars in the blending group. All other variables are constituents of ambiguously blended groups where a clear primary cannot be determined using available photometric and RV data.

\begin{figure}
\centering
\includegraphics[width=0.48\textwidth]{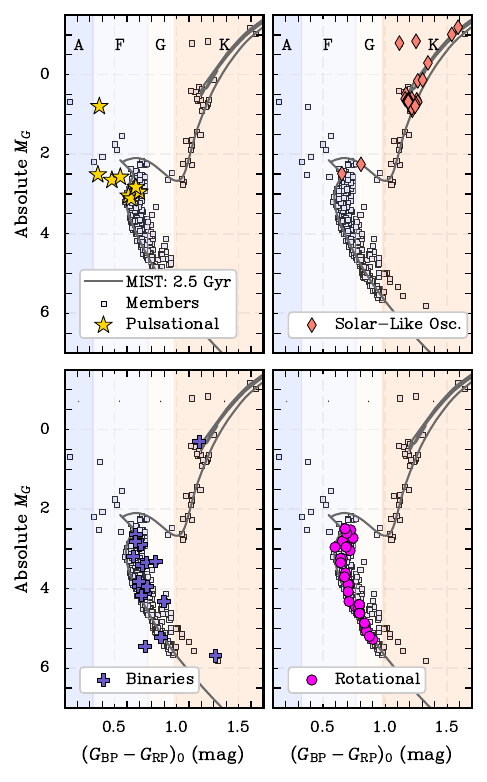}
\caption{Color-magnitude diagram showing the positions of our cluster variable sources, separated into four panels by variability class: pulsational, solar-like oscillators, binaries, and rotational variables. High-probability NGC\,6819 members ($P>0.9$) are overplotted as grey boxes. The solid curve is a 2.5\,Gyr MIST isochrone with [Fe/H] = $+0.05$\,dex \citep{Choi2016}. We annotate the top of the figure to indicate the range of spectral types, adding shaded panels that correspond to the hexadecimal color codes for a given spectral type  (corresponding to Luminosity class V and subclass 5), as reported by \cite{Harre2021}.}
\label{fig:cmdvar}
\end{figure}

In Figure~\ref{fig:cmdvar}, we illustrate the CMD positions of our cluster variables, with each panel broken into four distinct classes of variable types (pulsational variables, solar-like oscillators, binary variables, and rotational variables). 

Individual \textit{Gaia} GSP-Phot reddening estimates become unreliable for faint stars, introducing scatter that artificially broadens the cluster sequence \citep{Andrae2023}. 
Given that all cluster members lie at a common distance behind the same foreground dust, a uniform correction preserves the intrinsic tightness of the CMD and ensures consistent alignment with the cluster isochrone. 
We therefore apply a uniform reddening correction of $E(G_{\rm BP} - G_{\rm RP}) = 0.18$\,mag to all NGC\,6819 sources, adopting the median GSP-Phot value for bright (\textit{Kepler} magnitude
$<16.5$\,mag), high-parallax-precision (\texttt{parallax\_over\_error $> 11$}) cluster members.
For reference, we illustrate a Mesa Isochrones and Stellar Tracks (MIST) isochrone \citep{Paxton2011, Paxton2013, Paxton2015, Choi2016, Paxton2018} corresponding to the canonical cluster parameters of 2.5\,Gyr and [Fe/H] = 0.05\,dex.

\subsection{Rotation-Color Relation}\label{subsec:rotseq}

Of the 28 rotational variables in our catalog exhibiting periodic spot modulations, \totnewrot{} are new detections that were unreported by prior investigations.
More specifically, 22 of our rotational variables were not reported by \citet{Sanjayan2022}. Further, five of our rotational variables overlap with the rotation‐period sample produced by \citet{Meibom2015}, who also used four years of \textit{Kepler} data to calibrate gyrochronology (arriving at an age of 2.5\,Gyr). Like our analysis, this study targeted moderate period ranges and did not fully recover the slowest rotators among the faintest stars. For all five overlapping stars, our periods agree with those of \citet{Meibom2015} to within 10\%.

\begin{figure}
\centering
\includegraphics[width=0.48\textwidth]{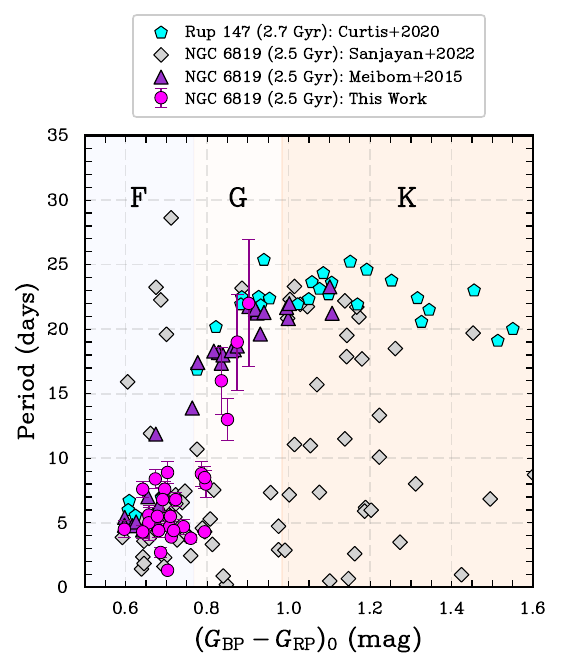}
    \caption{Rotation period versus dereddened \textit{Gaia} color for NGC\,6819 and Ruprecht\,147 (cyan; \citealt{Curtis2020}). Data sources for NGC\,6819 include this work (magenta), \citet{Meibom2015} (purple), and \citet{Sanjayan2022} (gray). Background shading indicates spectral type, as described in Figure~\ref{fig:cmdvar}.}
\label{fig:rotation}
\end{figure}

Figure~\ref{fig:rotation} illustrates the rotation period versus dereddened color for NGC\,6819 rotational variables, placing our catalog variables in the context of prior studies by \citet{Meibom2015} and \citet{Sanjayan2022}. 
For comparison, we include {benchmark stars from} the 2.7\,Gyr open cluster Ruprecht 147 \citep{Curtis2020}, which exhibits rotation periods consistent with spin-down evolution of stars in NGC\,6819 given the similar age.
Reddening corrections of $E(G_{\rm BP} - G_{\rm RP}) = 0.18$ and 0.15\,mag were applied to NGC\,6819 and Ruprecht\,147 \citep{2018A&A...619A.176B}, respectively.

We illustrate our rotational variables with period uncertainties, as described in Section~\ref{subsec:periodograms}, accounting for empirical variability from spot evolution, quarter-specific systematics, as well as theoretical resolution limits. 
Also described in Section~\ref{subsec:periodograms} is the fact that our periodogram analysis did not include periods longer than one-third the temporal baseline of a given quarter. Rare instances of larger reported periods arise when a harmonic was detected and, upon further inspection, the true period was found to be a multiplicative factor of the initial value (generally $2\times$). Consequently, the work of \citet{Meibom2015} and \citet{Sanjayan2022} reveal many more long-period variables that complement our sample.

In total, our sample contributes \totnewrot{} new rotation period measurements, increasing the number of known rotators in this 2.5\,Gyr benchmark cluster. 
The scatter in our sample is consistent with that observed in many coeval systems, reflecting both astrophysical and observational factors. 
For example, unresolved binaries (which we intentionally do not exclude) can appear as outliers below the gyrochronological sequence due to tidal spin-up. Additionally, differential rotation causes spots at different latitudes to trace different rotation rates, introducing scatter of $\sim$10--20\% in measured periods \citet{Reinhold_2015}. 
Spot evolution over baseline further contributes to quarter-to-quarter variability in period measurements. 

Notably, there is significant scatter in the \citet{Sanjayan2022} sample, which is likely driven by photometric blending in this crowded field.
In contrast, the \citet{Meibom2015} sample exhibits remarkably low scatter compared to other investigations of NGC\,6819 itself, including our analysis and the \citet{Sanjayan2022} results.
This may be reflective of their rigorous sample selection criteria, which included WOCS RV screening to exclude binaries and other potential contaminants. 
Readers interested in calibrating gyrochronological relations may apply additional membership and binarity cuts to our public catalog (such as stringent RUWE cuts to minimize contamination from unresolved binaries).

\subsubsection{Comparison with IRIS Rotational Variables}
\label{subsubsec:comparison}
The rotational variables in our catalog are largely complementary to the recent rotation period study of 
\citet{Sagynbayeva2025}, who identified 271 rotators in NGC\,6819 using Gaussian Process modeling of the IRIS light curves \citep{Colman2022}. Of their sample, only three stars overlap with the sources in our variable catalog (KIC 5112856, KIC 5024493, and KIC 5024534). All had reported periods that overlap with our results, agreeing to within 10\%.

We had independently flagged KIC\,5112856 and KIC\,5024534 as new detections but have since changed our designation.  We disagreed with the variable classification of KIC\,5112856, finding that the CMD position, periodicity, amplitude, and overall light curve morphology indicate it is likely a pulsational variable. 
Sources present in their catalog but absent from ours either fell below our detection thresholds or were removed during our  deblending analysis, which significantly pruned our initial sample of $\sim$550 cluster variables to the \totcat{} sources in the final catalog. 

For example, we found that the rotational variable KIC\,5024013 listed in \citet{Sagynbayeva2025} exhibited a broad, low-power periodogram with no peak clearly above the noise floor.
Another example is the KIC\,5024017, which we identified as a blend in our analysis of crowded regions. The limited overlap reflects the differing  sensitivities of the two approaches, as Gaussian Process methods are particularly effective at extracting low-amplitude, quasi-periodic rotational signals that lack a dominant periodogram peak \citep[e.g.,][]{Angus2018}, while we prioritized high-confidence detections using periodogram analysis and robust  deblending.

\section{Summary and Conclusions}\label{sec:summary}

We have applied image-subtraction and trend-filtering techniques to the data from \textit{Kepler} Quarters 1-16 superstamp of NGC\,6819, producing high-precision light curves for this crowded 2.5\,Gyr benchmark cluster and the surrounding region. Our principal results are as follows:

\begin{itemize}
    \item We extracted \totlcs{} light curves for \totstars{} unique sources, achieving a best 6.5-hr RMS of \bestppma{} at $K_{\mathrm{p}}=12$--12.5\,mag and \bestppmb{} at $K_{\mathrm{p}}=14$--14.5\,mag.
    
    \item Using \textit{Gaia} eDR3 astrometry, we identified 2,354 likely cluster members ($P_{\mathrm{mem}} > 0.5$) within 23\arcmin{} of the cluster center.
    
    \item We cataloged \totcat{} periodic variables, including 28 rotational variables, 25 solar-like oscillators, 18 binaries, five main-sequence pulsators, four pulsating blue stragglers, and seven miscellaneous variables.
    
    \item Of these, \totnewvar{} are newly identified, including \totnewrot{} new rotational variables.
\end{itemize}

We advise readers interested in calibrating gyrochronological relations using our data to apply stringent quality cuts to our public catalog, particularly RUWE constraints to minimize contamination from unresolved binaries.

\edit{The image-subtracted light curves presented in this work are available as a High Level Science Product (HLSP) at the Mikulski Archive for Space Telescopes (MAST) via \dataset[10.17909/g2w2-7965]{\doi{10.17909/g2w2-7965}}.}
\footnote{\url{https://archive.stsci.edu/hlsp/kplr-imsub/}}
The full cluster variable catalog is accessible via VizieR. This comprehensive dataset will enable studies of stellar variability, binary evolution, and angular-momentum evolution in middle-aged open clusters. Future work will focus on spectroscopic follow-up of newly discovered variables, dynamical modeling of the cluster binary population, and comparative analyses with other well-characterized clusters.

\begin{acknowledgments}
We thank J.~Auman, J.~Wallace, and I.~Colman for the insightful conversations related to this work.
We thank R.~Parsons for his generous assistance in getting our team server up and running. 
This work was catalyzed during the  Lamat REU program supported by NSF grant 1852393.
MSF gratefully acknowledges the generous support provided by NASA through Hubble Fellowship grant HST-HF2-51493.001-A awarded by the Space Telescope Science Institute, which is operated by the Association of Universities for Research in Astronomy, Inc., for NASA, under the contract NAS 5-26555.
Support for this research was provided by the Office of the Vice Chancellor for Research and Graduate Education at the University of Wisconsin--Madison with funding from the Wisconsin Alumni Research Foundation.
RLM acknowledges funding support from NSF AST-1714506 and the UW--Madison J.D. Fluno Family Distinguished Graduate Fellowship.
RDM acknowledges the support of the National Science Foundation through award NSF AST-1714506 and the Wisconsin Alumni Research Fund.

The data in our analysis were provided by the \textit{Kepler} mission, funded by the NASA Science Mission directorate. 
Photometric data were downloaded from the Barbara A.~Mikulski Archive for Space Telescopes (MAST). 
This research has made use of NASA's Astrophysics Data System Bibliographic Services. 
We made extensive use of the WEBDA database, operated at the Department of Theoretical Physics and Astrophysics of the Masaryk University. 
This work has made use of data from the European Space Agency (ESA) mission \emph{Gaia},\footnote{\url{https://www.cosmos.esa.int/gaia}} processed by the \emph{Gaia} Data Processing and Analysis Consortium (DPAC).\footnote{\url{https://www.cosmos.esa.int/web/gaia/dpac/consortium}} 
Funding for the DPAC has been provided by national institutions, in particular
the institutions participating in the \textit{Gaia} Multilateral Agreement.
This research has made use of NASA's Astrophysics Data System
Bibliographic Services.
This research has made use of the VizieR catalogue access tool, CDS, Strasbourg, France. The original description of the VizieR service was published in A\&AS 143, 23. 
\edit{This work made use of Astropy:\footnote{\url{http://www.astropy.org}} a community-developed core Python package and an ecosystem of tools and resources for astronomy \citep{astropy:2013, astropy2018, astropy:2022}.}
\end{acknowledgments}

\facilities{NASA-ADS Abstract Service; \textit{Gaia} \citep{GaiaDR3}; \textit{Kepler} \citep{Borucki2010}; Mikulski Archive for Space Telescopes \citep{MAST};
WIYN/Hydra}

\software{
    \texttt{Astrobase} \citep{Bhatti2018}; 
    \texttt{Astropy} \citep{astropy:2013, astropy2018, astropy:2022};
    \texttt{astroquery} \citep{astroquery};     
    \texttt{FITSH} \citep{pal};    \texttt{LcTools}\footnote{\url{https://sites.google.com/a/lctools.net/lctools}} \citep{schmitt2019lctools,schmitt2021lctools};
    \texttt{Lightkurve}  \citep{lightkurve};
    \texttt{Matplotlib} \citep{matplotlib};
    \texttt{MESA} \citep{Paxton2011,Paxton2013,Paxton2015,Paxton2018}, 
    \texttt{MIST} \citep{Choi2016};
    \edit{\texttt{NumPy} \citep{2020NumPy-Array}}; 
    \texttt{Pandas} \citep{Pandas};
    \edit{\texttt{SciPy} \citep{2020SciPy-NMeth}};
    \edit{\texttt{smplotlib} \citep{smplotlib}};
    \texttt{VARTOOLS} \citep{vartools};
    VizieR catalogue access tool \citep{Vizier}
} 

\pagebreak
\appendix

\startlongtable
\begin{deluxetable*}{clcccccccccc}
\tablecaption{ \label{bigtable}}
\tablehead{
  \colhead{No.}                 &
  \colhead{\textit{Gaia} DR3 ID} & 
  \colhead{KIC ID}              &
  \colhead{\textit{$M_G$}}     & 
  \colhead{$P$}                 & 
  \colhead{$\Delta$mag}         &
  \colhead{SNR}                 &
  \colhead{Class}               & 
  \colhead{New}                 &
  \colhead{Blend}               & 
  \colhead{$P_{\mathrm{mem}}$}  & 
  \colhead{Provenance}         \\ 
  \colhead{}                    &
  \colhead{}                    & 
  \colhead{}                    &
  \colhead{(mag)}               & 
  \colhead{(days)}              & 
  \colhead{(mmag)}              & 
  \colhead{}                    & 
  \colhead{}                    & 
  \colhead{}                    & 
  \colhead{}                    &
  \colhead{}                    &
  \colhead{}
}
\startdata
1 & 2076298417671932800 & 5024455 & 14.752 & $1.16 \pm 0.01$ & 2.6 & 10.3 & 1a & [] & 0 & 0.996 & W06, Th15 \\
2 & 2076299482823088000\tablenotemark{a} & 5024084 & 14.674 & $2.04 \pm 0.04$ & 22.4 & 98.2 & 1a & [] & 0 & 0.994 & W06, MQ14 \\
3 & 2076299826420542080\tablenotemark{a} & 5024468 & 12.898 & $0.08 \pm 0.01$ & 7.3 & 176.7 & 1a & [] & 0 & 0.996 & S22, G23 \\
4 & 2076393594145246080 & 5113357 & 14.613 & $0.06 \pm 0.01$ & 2.5 & 150.9 & 1a & [3] & 0 & 0.996 & S22 \\
5 & 2076299792060826624 & 5024615 & 15.146 & $0.30 \pm 0.01$ & 12.3 & 70.2 & 1b & [2] & 0 & 0.996 & W06, Ch20 \\
6 & 2076299895139969536 & 5024202 & 15.22 & $2.89 \pm 0.09$ & 1.2 & 82.9 & 1b & [2] & 1 & 0.91 & W06, MQ14 \\
7 & 2076393113108862080 & 5025288 & 15.05 & $1.39 \pm 0.02$ & 2.3 & 128.3 & 1b & [2] & 0 & 0.95 & R23, G23 \\
8 & 2076394040821761152 & 5112856 & 14.93 & $8.0 \pm 0.7$ & 2.4 & 12.4 & 1b & [2] & 1 & 0.93 & Sg25 \\
9 & 2076488049065148288 & \nodata & 15.98 & $1.90 \pm 0.04$ & 6.2 & 71.3 & 1b & [2] & 0 & 0.74 & S22 \\
10 & 2076299757701068800 & 5024600 & 16.221 & $3.8 \pm 0.1$ & 6.5 & 68.1 & 2a & [2] & 1 & 0.99 & S22 \\
11 & 2076299929499734912 & 5024292 & 15.001 & $2.3 \pm 0.1$ & 31.2 & 10.4 & 2a & [] & 0 & 0.6 & Th15, Lu17 \\
12 & 2076300066938729856\tablenotemark{a} & 5112574 & 16.045 & $1.60 \pm 0.03$ & 7.4 & 24.5 & 2a & [2] & 0 & 0.994 & S22, G23 \\
\rowcolor{yellow!25} 13 & 2076300101294350976 & \nodata & 17.554 & $1.04 \pm 0.02$ & 50.7 & 15.8 & 2a & [1] & 1 & 0.91 &  \\
\rowcolor{yellow!25} 14 & 2076300101298445824 & 5112474 & 17.317 & $1.05 \pm 0.02$ & 40.7 & 11.2 & 2a & [1] & 1 & 0.85 &  \\
15 & 2076300101299126912\tablenotemark{a} & \nodata & 14.742 & $4.9 \pm 0.4$ & 1.9 & 37.9 & 2a & [2] & 1 & 0.89 & S22 \\
16 & 2076389608415499392\tablenotemark{a} & 5025105 & 15.286 & $1.58 \pm 0.06$ & 3.7 & 14.3 & 2a & [] & 0 & 0.89 & S22, G23 \\
17 & 2076393692914036608\tablenotemark{a} & 5112741 & 12.408 & $17.6 \pm 0.8$ & 22.9 & 13.2 & 2a & [] & 0 & 0.995 & W06, S22 \\
18 & 2076487121349914752\tablenotemark{a} & 5023948 & 14.991 & $3.6 \pm 0.5$ & 14.0 & 8.4 & 2a & [] & 0 & 0.9 & Th15, Lu17 \\
19 & 2076299036142588032\tablenotemark{a} & 5024760 & 16.043 & $2.47 \pm 0.06$ & 4.7 & 42.0 & 2b & [2] & 0 & 0.93 & S22 \\
20 & 2076392838230907392\tablenotemark{a} & 5024988 & 15.502 & $5.6 \pm 0.4$ & 5.7 & 11.4 & 2b & [2] & 0 & 0.94 & S22, G23 \\
21 & 2076393044389355008\tablenotemark{a} & 5113053 & 15.445 & $3.2 \pm 0.1$ & 5.6 & 36.1 & 2b & [2] & 0 & 0.77 & Th15, Lu17 \\
22 & 2076393731584598400 & 5112843 & 15.422 & $0.17 \pm 0.01$ & 121.7 & 148.7 & 2b & [] & 0 & 0.996 & S22, G23 \\
23 & 2076581644988574592 & 5112536 & 16.266 & $0.75 \pm 0.01$ & 56.4 & 165.4 & 2b & [] & 1 & 0.93 & S22 \\
24 & 2076581713708050176 & 5112407 & 16.415 & $0.18 \pm 0.01$ & 259.1 & 160.8 & 2b & [] & 0 & 0.92 & S22, G23 \\
25 & 2076393903382776576\tablenotemark{a} & 5112699 & 15.925 & $2.28 \pm 0.08$ & 5.2 & 21.9 & 2c & [] & 0 & 0.95 & J20, S22 \\
26 & 2076394105234042496 & 5112759 & 17.788 & $0.26 \pm 0.01$ & 64.9 & 60.2 & 2c & [] & 0 & 0.87 & S22, G23 \\
27 & 2076581782427559808 & 5112708 & 14.915 & $1.72 \pm 0.03$ & 12.8 & 12.7 & 2c & [2] & 0 & 0.91 & S22, G23 \\
28 & 2076298795628342144 & 5024493 & 14.806 & $4.7 \pm 0.2$ & 1.1 & 13.0 & 3 & [] & 0 & 0.93 & Sa21, G23 \\
\rowcolor{yellow!25} 29 & 2076298829988103424 & 5024595 & 16.555 & $3.8 \pm 0.3$ & 4.2 & 6.7 & 3 & [1] & 0 & 0.94 &  \\
\rowcolor{yellow!25} 30 & 2076299036146572288 & 5024723 & 14.904 & $5.6 \pm 0.6$ & 0.8 & 7.2 & 3 & [1] & 0 & 0.74 &  \\
\rowcolor{yellow!25} 31 & 2076299139225682944 & 5024040 & 17.167 & $13 \pm 2$ & 127.5 & 17.1 & 3 & [1] & 0 & 0.9 &  \\
32 & 2076299306716879744 & 5024280 & 16.966 & $16 \pm 3$ & 7.2 & 12.6 & 3 & [] & 0 & 0.94 & Me15 \\
33 & 2076299723341302272 & 5024403 & 15.704 & $4.4 \pm 0.6$ & 1.9 & 11.4 & 3 & [] & 0 & 0.93 & S22 \\
\rowcolor{yellow!25} 34 & 2076299723341311616 & 5024453 & 15.341 & $5 \pm 1$ & 4.7 & 6.7 & 3 & [1] & 0 & 0.91 &  \\
\rowcolor{yellow!25} 35 & 2076299757701063936 & 5024598 & 15.989 & $3.9 \pm 0.2$ & 1.7 & 14.4 & 3 & [1] & 0 & 0.985 &  \\
\rowcolor{yellow!25} 36 & 2076299792060823040 & 5024663 & 15.043 & $7.6 \pm 0.6$ & 1.5 & 8.9 & 3 & [1] & 1 & 0.96 &  \\
\rowcolor{yellow!25} 37 & 2076299792060823680 & 5024653 & 15.338 & $7.6 \pm 0.6$ & 1.8 & 5.9 & 3 & [1] & 1 & 0.95 &  \\
\rowcolor{yellow!25} 38 & 2076299826420530048 & \nodata & 14.825 & $4.7 \pm 0.6$ & 1.2 & 8.4 & 3 & [1] & 0 & 0.94 &  \\
39 & 2076299860780301056 & 5024534 & 16.722 & $8 \pm 1$ & 74.1 & 9.3 & 3 & [] & 0 & 0.94 & Sg25 \\
\rowcolor{yellow!25} 40 & 2076299895139972992 & 5024250 & 15.99 & $1.32 \pm 0.02$ & 8.9 & 60.1 & 3 & [1] & 0 & 0.89 &  \\
\rowcolor{yellow!25} 41 & 2076299929495609088 & \nodata & 15.803 & $8.4 \pm 0.7$ & 10.0 & 6.7 & 3 & [1] & 0 & 0.95 &  \\
42 & 2076299929499804288 & 5024287 & 14.623 & $6.8 \pm 0.5$ & 1.4 & 9.7 & 3 & [] & 0 & 0.99 & S22 \\
43 & 2076300066938738176 & 5112630 & 15.141 & $4.4 \pm 0.2$ & 2.7 & 51.9 & 3 & [] & 0 & 0.994 & S22, G23 \\
44 & 2076300135658207232 & 5112512 & 16.422 & $5.5 \pm 0.4$ & 8.1 & 9.8 & 3 & [] & 0 & 0.78 & S22 \\
\rowcolor{yellow!25} 45 & 2076393697224860672 & 5024638 & 16.176 & $8.9 \pm 0.8$ & 113.2 & 10.4 & 3 & [1] & 0 & 0.925 &  \\
\rowcolor{yellow!25} 46 & 2076393731584100736 & 5112863 & 15.054 & $2.7 \pm 0.4$ & 0.8 & 7.6 & 3 & [1] & 0 & 0.992 &  \\
47 & 2076393899075929472 & 5112711 & 16.589 & $8.8 \pm 0.9$ & 24.4 & 12.3 & 3 & [] & 0 & 0.92 & S22 \\
\rowcolor{yellow!25} 48 & 2076393899075929728 & 5112726 & 16.498 & $8.5 \pm 0.8$ & 6.4 & 29.3 & 3 & [1] & 0 & 0.94 &  \\
49 & 2076393903382760576 & 5112677 & 14.707 & $6.8 \pm 0.9$ & 1.3 & 31.4 & 3 & [] & 0 & 0.993 & S22 \\
50 & 2076394006454008320 & 5112675 & 16.72 & $4.3 \pm 0.2$ & 8.0 & 36.0 & 3 & [] & 0 & 0.92 & S22 \\
51 & 2076487185762452096 & 5023899 & 15.056 & $4.5 \pm 0.7$ & 1.1 & 7.1 & 3 & [] & 0 & 0.95 & Me15, G23 \\
\rowcolor{yellow!25} 52 & 2076487529359867392 & 5023909 & 17.374 & $22 \pm 5$ & 13.2 & 52.6 & 3 & [1] & 0 & 0.93 &  \\
53 & 2076488044756008576 & 5112268 & 17.299 & $19 \pm 4$ & 11.6 & 30.3 & 3 & [] & 0 & 0.9 & Me15, Sa21 \\
54 & 2076488942416130688 & 5112104 & 14.593 & $5.5 \pm 0.4$ & 1.1 & 26.3 & 3 & [] & 0 & 0.94 & W06, MQ14 \\
55 & 2076581748067801088 & 5112499 & 15.448 & $4.3 \pm 0.5$ & 0.9 & 7.0 & 3 & [] & 0 & 0.91 & Me15, G23 \\
\rowcolor{yellow!25} 56 & 2076299719037114496 & 5024323 & 17.365 & $4.1 \pm 0.8$ & 13.4 & 6.8 & 4 & [1] & 0 & 0.87 &  \\
\rowcolor{yellow!25} 57 & 2076392872590611328 & \nodata & 16.231 & $3.5 \pm 0.1$ & 5.5 & 42.5 & 4 & [1] & 1 & 0.91 &  \\
58 & 2076392872590614784 & 5024751 & 17.904 & $3.5 \pm 0.1$ & 28.4 & 47.0 & 4 & [] & 1 & 0.9 & S22, G23 \\
\rowcolor{yellow!25} 59 & 2076393624197664768 & 5024683 & 17.047 & $3.4 \pm 0.1$ & 17.0 & 31.0 & 4 & [1] & 1 & 0.85 &  \\
60 & 2076393628504856448 & 5024697 & 16.109 & $3.4 \pm 0.1$ & 7.8 & 36.1 & 4 & [] & 1 & 0.92 & S22 \\
61 & 2076393933435330688 & 5112777 & 17.445 & $6.1 \pm 0.5$ & 16.2 & 9.5 & 4 & [] & 0 & 0.82 & S22 \\
\rowcolor{yellow!25} 62 & 2076487258788899456 & 5024016 & 15.0 & $0.79 \pm 0.01$ & 1.1 & 48.6 & 4 & [1] & 1 & 0.96 &  \\
\rowcolor{yellow!25} 63 & 2076299448463334144 & 5023953 & 12.589 & $0.8 \pm 0.6$ & 0.2 & 7.9 & 5 & [1] & 0 & 0.989 &  \\
64 & 2076299792060819328 & 5024583 & 12.732 & $1.3 \pm 0.3$ & 0.4 & 7.1 & 5 & [2] & 0 & 0.995 & S22 \\
\rowcolor{yellow!25} 65 & 2076299826420588416 & 5024420 & 14.355 & $3.0 \pm 0.3$ & 3.4 & 8.1 & 5 & [1] & 0 & 0.88 &  \\
66 & 2076299856476070784 & 5024511 & 12.256 & $0.55 \pm 0.07$ & 0.4 & 31.7 & 5 & [3] & 1 & 0.995 & S22 \\
\rowcolor{yellow!25} 67 & 2076299860780295936 & 5024517 & 12.995 & $0.56 \pm 0.07$ & 0.7 & 31.6 & 5 & [1] & 1 & 0.991 &  \\
\rowcolor{yellow!25} 68 & 2076299993911496320 & 5024268 & 11.306 & $0.5 \pm 0.1$ & 0.2 & 8.9 & 5 & [1] & 0 & 0.996 &  \\
69 & 2076299998219216256 & 5024272 & 11.26 & $1.7 \pm 0.9$ & 0.5 & 14.3 & 5 & [] & 0 & 0.994 & G23 \\
70 & 2076300101294387584 & 5112491 & 12.711 & $1.38 \pm 0.08$ & 0.4 & 7.3 & 5 & [] & 0 & 0.991 &  \\
71 & 2076300101298449280 & 5112467 & 12.757 & $1.3 \pm 0.1$ & 0.4 & 5.9 & 5 & [3] & 0 & 0.994 &  \\
\rowcolor{yellow!25} 72 & 2076392528993233024 & 5024967 & 12.793 & $0.27 \pm 0.03$ & 0.1 & 17.9 & 5 & [1] & 0 & 0.64 &  \\
\rowcolor{yellow!25} 73 & 2076393010029609984 & 5113041 & 12.789 & $0.32 \pm 0.03$ & 0.3 & 34.8 & 5 & [1] & 0 & 0.987 &  \\
74 & 2076393624194561152 & 5024750 & 11.806 & $1.0 \pm 0.2$ & 0.7 & 15.8 & 5 & [] & 0 & 0.985 & S22 \\
75 & 2076393658554297600 & 5112880 & 12.235 & $0.44 \pm 0.03$ & 0.4 & 27.1 & 5 & [] & 0 & 0.994 &  \\
76 & 2076393662864626176\tablenotemark{a} & 5112950 & 12.706 & $0.27 \pm 0.02$ & 0.2 & 11.0 & 5 & [] & 0 & 0.99 & G23 \\
77 & 2076393765943846656 & 5112938 & 12.762 & $1.2 \pm 0.2$ & 0.3 & 7.0 & 5 & [] & 0 & 0.996 & S22 \\
78 & 2076393869023076608 & 5112948 & 12.833 & $0.27 \pm 0.02$ & 0.2 & 15.7 & 5 & [] & 0 & 0.996 & S22 \\
79 & 2076393903374692992 & 5112666 & 14.59 & $7.1 \pm 0.6$ & 1.5 & 6.5 & 5 & [] & 0 & 0.77 &  \\
80 & 2076393937742523904 & 5112730 & 12.719 & $0.27 \pm 0.01$ & 0.2 & 11.3 & 5 & [] & 0 & 0.94 & S22 \\
81 & 2076394659297111552 & 5113061 & 11.088 & $2.5 \pm 0.3$ & 1.4 & 48.2 & 5 & [] & 0 & 0.99 & S22 \\
82 & 2076487224429134592\tablenotemark{a} & 5023931 & 12.94 & $1.1 \pm 0.9$ & 0.3 & 20.5 & 5 & [] & 0 & 0.983 &  \\
83 & 2076487838597288320 & 5112373 & 12.755 & $0.26 \pm 0.02$ & 0.2 & 11.7 & 5 & [] & 0 & 0.995 & S22 \\
84 & 2076487872957025408 & 5112387 & 12.776 & $0.3 \pm 0.1$ & 0.2 & 19.6 & 5 & [] & 0 & 0.987 & S22, G23 \\
85 & 2076488049062931328 & 5112288 & 12.778 & $0.27 \pm 0.03$ & 0.2 & 10.0 & 5 & [] & 0 & 0.991 &  \\
86 & 2076581816787312896 & 5112744 & 12.89 & $0.28 \pm 0.03$ & 0.2 & 43.2 & 5 & [] & 0 & 0.996 & S22 \\
87 & 2076583534763920896 & 5200056 & 10.906 & $3.4 \pm 0.8$ & 6.1 & 18.2 & 5 & [] & 0 & 0.996 & G23 \\
\enddata
\tablecomments{Subset of columns from our variable catalog. Columns include \textbf{No.}: index number; \textbf{Gaia DR3 ID}: \textit{Gaia} DR3 source identifier---the superscript $a$ indicates that a WOCS radial velocity orbital solution is available for this target and aided in the classification; \textbf{KIC ID}: Kepler Input Catalog identifier (when available); \textbf{$\mathbf{M_G}$}: \textit{Gaia} apparent magnitude; \textbf{$\mathbf{P}$}: median photometric period measured across all available quarter and its associated error; \textbf{$\Delta$mag}: peak-to-peak variability amplitude from the phase-folded light curve; \textbf{SNR}: maximum signal-to-noise ratio across quarters, defined as the ratio of the maximum Lomb-Scargle periodogram power to the MAD of the power spectrum; \textbf{Class}: variable-type classification (1a = blue straggler pulsator, 1b = pulsator of uncertain type (likely $\gamma$ Doradus and $\delta$ Scuti variables), 2a = detached eclipsing binary, 2b = contact binary, 2c = semi-detached/EB binary, 3 = rotational variable, 4 = variable of indeterminate type, 5 = solar-like oscillator); \textbf{New}: comma-separated list of novel contributions: 1 = new detection (not in previous NGC 6819 variable catalogs), 2 = new classification (previously known variable with updated type), 3 = new/refined period (improved period measurement), [] = previously cataloged variables with no updates; \textbf{Blend}: blended signal indicator (1 = yes, 0 = no); \textbf{$\mathbf{P_{\mathrm{mem}}}$}: cluster membership probability; and \textbf{Provenance} indicates the two earliest references noting variability for this target 
(W06: \citealt{Watson2006}; 
MQ14: \citealt{Mcquillan2014};
Me15: \citealt{Meibom2015};
Th15: \citealt{Thompson2015};
Lu17: \citealt{Lurie2017}; 
Ch20: \citealt{Chen2020}; 
J20: \citealt{Jayasinghe2020};
Sa21: \citealt{Santos2021};
S22: \citealt{Sanjayan2022}; 
G23: \citealt{Gavras2023}; 
R23: \citealt{Reinhold2023}; 
Sg25: \citealt{Sagynbayeva2025}). 
Rows highlighted in yellow denote new detections (those carrying a New column code of [1]).
The full catalog is available via VizieR and contains additional information beyond the column subset shown here. All magnitudes are drawn from \textit{Gaia} DR3, providing a uniformly calibrated photometric system with millimagnitude precision. This ensures consistent coverage for all \textit{Kepler} superstamp sources and aligns with our membership analysis. For these targets, the \textit{Kepler} and \textit{Gaia} $G$ magnitudes typically differ by only ${\sim}0.1$\,mag, so adopting $G$ has a negligible effect on the overall brightness scale.}
\end{deluxetable*}

\bibliography{bibliography.bib}

@ARTICLE{Platais2013,
       author = {{Platais}, Imants and {Gosnell}, Natalie M. and {Meibom}, S{\o}ren and {Kozhurina-Platais}, Vera and {Bellini}, Andrea and {Veillet}, Christian and {Burkhead}, Martin S.},
        title = "{WIYN Open Cluster Study. LV. Astrometry and Membership in NGC 6819}",
      journal = {\aj},
         year = 2013,
        month = aug,
       volume = {146},
       number = {2},
          eid = {43},
        pages = {43},
          doi = {10.1088/0004-6256/146/2/43},
       adsurl = {https://ui.adsabs.harvard.edu/abs/2013AJ....146...43P}
}

@ARTICLE{Ak2016,
       author = {{Ak}, T. and {Bostanc{\i}}, Z.~F. and {Yontan}, T. and {Bilir}, S. and {G{\"u}ver}, T. and {Ak}, S. and {{\"U}rg{\"u}p}, H. and {Paunzen}, E.},
        title = "{CCD UBV photometry of the open cluster NGC 6819}",
      journal = {\apss},
         year = 2016,
        month = apr,
       volume = {361},
          eid = {126},
        pages = {126},
          doi = {10.1007/s10509-016-2707-2},
archivePrefix = {arXiv},
       eprint = {1603.00474},
 primaryClass = {astro-ph.GA},
       adsurl = {https://ui.adsabs.harvard.edu/abs/2016Ap&SS.361..126A}
}

@ARTICLE{Barnes2007,
       author = {{Barnes}, Sydney A.},
        title = "{Ages for Illustrative Field Stars Using Gyrochronology: Viability, Limitations, and Errors}",
      journal = {\apj},
         year = 2007,
        month = nov,
       volume = {669},
       number = {2},
        pages = {1167-1189},
          doi = {10.1086/519295},
archivePrefix = {arXiv},
       eprint = {0704.3068},
 primaryClass = {astro-ph},
       adsurl = {https://ui.adsabs.harvard.edu/abs/2007ApJ...669.1167B}
}

@ARTICLE{Choi2016,
       author = {{Choi}, Jieun and {Dotter}, Aaron and {Conroy}, Charlie and {Cantiello}, Matteo and {Paxton}, Bill and {Johnson}, Benjamin D.},
        title = "{Mesa Isochrones and Stellar Tracks (MIST). I. Solar-scaled Models}",
      journal = {\apj},
         year = 2016,
        month = jun,
       volume = {823},
       number = {2},
          eid = {102},
        pages = {102},
          doi = {10.3847/0004-637X/823/2/102},
archivePrefix = {arXiv},
       eprint = {1604.08592},
 primaryClass = {astro-ph.SR},
       adsurl = {https://ui.adsabs.harvard.edu/abs/2016ApJ...823..102C}
}

@ARTICLE{Harre2021,
       author = {{Harre}, Jan-Vincent and {Heller}, Ren{\'e}},
        title = "{Digital color codes of stars}",
      journal = {Astronomische Nachrichten},
         year = 2021,
        month = mar,
       volume = {342},
       number = {3},
        pages = {578-587},
          doi = {10.1002/asna.202113868},
archivePrefix = {arXiv},
       eprint = {2101.06254},
 primaryClass = {astro-ph.SR},
       adsurl = {https://ui.adsabs.harvard.edu/abs/2021AN....342..578H}
}

@ARTICLE{Kovacs2002,
       author = {{Kov{\'a}cs}, G. and {Zucker}, S. and {Mazeh}, T.},
        title = "{A box-fitting algorithm in the search for periodic transits}",
      journal = {\aap},
         year = 2002,
        month = aug,
       volume = {391},
        pages = {369-377},
          doi = {10.1051/0004-6361:20020802},
archivePrefix = {arXiv},
       eprint = {astro-ph/0206099},
 primaryClass = {astro-ph},
       adsurl = {https://ui.adsabs.harvard.edu/abs/2002A&A...391..369K}
}

@ARTICLE{Paxton2011,
       author = {{Paxton}, Bill and {Bildsten}, Lars and {Dotter}, Aaron and {Herwig}, Falk and {Lesaffre}, Pierre and {Timmes}, Frank},
        title = "{Modules for Experiments in Stellar Astrophysics (MESA)}",
      journal = {\apjs},
         year = 2011,
        month = jan,
       volume = {192},
       number = {1},
          eid = {3},
        pages = {3},
          doi = {10.1088/0067-0049/192/1/3},
archivePrefix = {arXiv},
       eprint = {1009.1622},
 primaryClass = {astro-ph.SR},
       adsurl = {https://ui.adsabs.harvard.edu/abs/2011ApJS..192....3P}
}

@ARTICLE{Paxton2013,
       author = {{Paxton}, Bill and {Cantiello}, Matteo and {Arras}, Phil and {Bildsten}, Lars and {Brown}, Edward F. and {Dotter}, Aaron and {Mankovich}, Christopher and {Montgomery}, M.~H. and {Stello}, Dennis and {Timmes}, F.~X. and {Townsend}, Richard},
        title = "{Modules for Experiments in Stellar Astrophysics (MESA): Planets, Oscillations, Rotation, and Massive Stars}",
      journal = {\apjs},
         year = 2013,
        month = sep,
       volume = {208},
       number = {1},
          eid = {4},
        pages = {4},
          doi = {10.1088/0067-0049/208/1/4},
archivePrefix = {arXiv},
       eprint = {1301.0319},
 primaryClass = {astro-ph.SR},
       adsurl = {https://ui.adsabs.harvard.edu/abs/2013ApJS..208....4P}
}

@ARTICLE{Paxton2015,
       author = {{Paxton}, Bill and {Marchant}, Pablo and {Schwab}, Josiah and {Bauer}, Evan B. and {Bildsten}, Lars and {Cantiello}, Matteo and {Dessart}, Luc and {Farmer}, R. and {Hu}, H. and {Langer}, N. and {Townsend}, R.~H.~D. and {Townsley}, Dean M. and {Timmes}, F.~X.},
        title = "{Modules for Experiments in Stellar Astrophysics (MESA): Binaries, Pulsations, and Explosions}",
      journal = {\apjs},
         year = 2015,
        month = sep,
       volume = {220},
       number = {1},
          eid = {15},
        pages = {15},
          doi = {10.1088/0067-0049/220/1/15},
archivePrefix = {arXiv},
       eprint = {1506.03146},
 primaryClass = {astro-ph.SR},
       adsurl = {https://ui.adsabs.harvard.edu/abs/2015ApJS..220...15P}
}

@ARTICLE{Paxton2018,
       author = {{Paxton}, Bill and {Schwab}, Josiah and {Bauer}, Evan B. and {Bildsten}, Lars and {Blinnikov}, Sergei and {Duffell}, Paul and {Farmer}, R. and {Goldberg}, Jared A. and {Marchant}, Pablo and {Sorokina}, Elena and {Thoul}, Anne and {Townsend}, Richard H.~D. and {Timmes}, F.~X.},
        title = "{Modules for Experiments in Stellar Astrophysics (MESA): Convective Boundaries, Element Diffusion, and Massive Star Explosions}",
      journal = {\apjs},
         year = 2018,
        month = feb,
       volume = {234},
       number = {2},
          eid = {34},
        pages = {34},
          doi = {10.3847/1538-4365/aaa5a8},
archivePrefix = {arXiv},
       eprint = {1710.08424},
 primaryClass = {astro-ph.SR},
       adsurl = {https://ui.adsabs.harvard.edu/abs/2018ApJS..234...34P}
}

@ARTICLE{Schmitt2016,
       author = {{Schmitt}, Joseph R. and {Tokovinin}, Andrei and {Wang}, Ji and {Fischer}, Debra A. and {Kristiansen}, Martti H. and {LaCourse}, Daryll M. and {Gagliano}, Robert and {Tan}, Arvin Joseff V. and {Schwengeler}, Hans Martin and {Omohundro}, Mark R. and {Venner}, Alexander and {Terentev}, Ivan and {Schmitt}, Allan R. and {Jacobs}, Thomas L. and {Winarski}, Troy and {Sejpka}, Johann and {Jek}, Kian J. and {Boyajian}, Tabetha S. and {Brewer}, John M. and {Ishikawa}, Sascha T. and {Lintott}, Chris and {Lynn}, Stuart and {Schawinski}, Kevin and {Schwamb}, Megan E. and {Weiksnar}, Alex},
        title = "{Planet Hunters. X. Searching for Nearby Neighbors of 75 Planet and Eclipsing Binary Candidates from the K2 Kepler extended mission}",
      journal = {\aj},
         year = 2016,
        month = jun,
       volume = {151},
       number = {6},
          eid = {159},
        pages = {159},
          doi = {10.3847/0004-6256/151/6/159},
archivePrefix = {arXiv},
       eprint = {1603.06945},
 primaryClass = {astro-ph.EP},
       adsurl = {https://ui.adsabs.harvard.edu/abs/2016AJ....151..159S}
}

@ARTICLE{Mann2020,
       author = {{Mann}, Andrew W. and {Johnson}, Marshall C. and {Vanderburg}, Andrew and {Kraus}, Adam L. and {Rizzuto}, Aaron C. and {Wood}, Mackenna L. and {Bush}, Jonathan L. and {Rockcliffe}, Keighley and {Newton}, Elisabeth R. and {Latham}, David W. and {Mamajek}, Eric E. and {Zhou}, George and {Quinn}, Samuel N. and {Thao}, Pa Chia and {Benatti}, Serena and {Cosentino}, Rosario and {Desidera}, Silvano and {Harutyunyan}, Avet and {Lovis}, Christophe and {Mortier}, Annelies and {Pepe}, Francesco A. and {Poretti}, Ennio and {Wilson}, Thomas G. and {Kristiansen}, Martti H. and {Gagliano}, Robert and {Jacobs}, Thomas and {LaCourse}, Daryll M. and {Omohundro}, Mark and {Schwengeler}, Hans Martin and {Terentev}, Ivan A. and {Kane}, Stephen R. and {Hill}, Michelle L. and {Rabus}, Markus and {Esquerdo}, Gilbert A. and {Berlind}, Perry and {Collins}, Karen A. and {Murawski}, Gabriel and {Sallam}, Nezar Hazam and {Aitken}, Michael M. and {Massey}, Bob and {Ricker}, George R. and {Vanderspek}, Roland and {Seager}, Sara and {Winn}, Joshua N. and {Jenkins}, Jon M. and {Barclay}, Thomas and {Caldwell}, Douglas A. and {Dragomir}, Diana and {Doty}, John P. and {Glidden}, Ana and {Tenenbaum}, Peter and {Torres}, Guillermo and {Twicken}, Joseph D. and {Villanueva}, Steven, Jr.},
        title = "{TESS Hunt for Young and Maturing Exoplanets (THYME). III. A Two-planet System in the 400 Myr Ursa Major Group}",
      journal = {\aj},
         year = 2020,
        month = oct,
       volume = {160},
       number = {4},
          eid = {179},
        pages = {179},
          doi = {10.3847/1538-3881/abae64},
archivePrefix = {arXiv},
       eprint = {2005.00047},
 primaryClass = {astro-ph.EP},
       adsurl = {https://ui.adsabs.harvard.edu/abs/2020AJ....160..179M}
}

@ARTICLE{Powell2021,
       author = {{Powell}, Brian P. and {Kostov}, Veselin B. and {Rappaport}, Saul A. and {Borkovits}, Tam{\'a}s and {Zasche}, Petr and {Tokovinin}, Andrei and {Kruse}, Ethan and {Latham}, David W. and {Montet}, Benjamin T. and {Jensen}, Eric L.~N. and {Jayaraman}, Rahul and {Collins}, Karen A. and {Ma{\v{s}}ek}, Martin and {Hellier}, Coel and {Evans}, Phil and {Tan}, Thiam-Guan and {Schlieder}, Joshua E. and {Torres}, Guillermo and {Smale}, Alan P. and {Friedman}, Adam H. and {Barclay}, Thomas and {Gagliano}, Robert and {Quintana}, Elisa V. and {Jacobs}, Thomas L. and {Gilbert}, Emily A. and {Kristiansen}, Martti H. and {Col{\'o}n}, Knicole D. and {LaCourse}, Daryll M. and {Olmschenk}, Greg and {Omohundro}, Mark and {Schnittman}, Jeremy D. and {Schwengeler}, Hans M. and {Barry}, Richard K. and {Terentev}, Ivan A. and {Boyd}, Patricia and {Schmitt}, Allan R. and {Quinn}, Samuel N. and {Vanderburg}, Andrew and {Palle}, Enric and {Armstrong}, James and {Ricker}, George R. and {Vanderspek}, Roland and {Seager}, S. and {Winn}, Joshua N. and {Jenkins}, Jon M. and {Caldwell}, Douglas A. and {Wohler}, Bill and {Shiao}, Bernie and {Burke}, Christopher J. and {Daylan}, Tansu and {Villase{\~n}or}, Joel},
        title = "{TIC 168789840: A Sextuply Eclipsing Sextuple Star System}",
      journal = {\aj},
         year = 2021,
        month = apr,
       volume = {161},
       number = {4},
          eid = {162},
        pages = {162},
          doi = {10.3847/1538-3881/abddb5},
archivePrefix = {arXiv},
       eprint = {2101.03433},
 primaryClass = {astro-ph.SR},
       adsurl = {https://ui.adsabs.harvard.edu/abs/2021AJ....161..162P}
}

@misc{schmitt2019lctools,
      title={LcTools: A Windows-Based Software System for Finding and Recording Signals in Lightcurves from NASA Space Missions}, 
      author={Allan R. Schmitt and Joel D. Hartman and David M. Kipping},
      year={2019},
      eprint={1910.08034},
      archivePrefix={arXiv},
      primaryClass={astro-ph.IM}
}

@misc{schmitt2021lctools,
      title={LcTools II: The QuickFind Method for Finding Signals and Associated TTVs in Light Curves from NASA Space Missions}, 
      author={Allan Schmitt and Andrew Vanderburg},
      year={2021},
      eprint={2103.10285},
      archivePrefix={arXiv},
      primaryClass={astro-ph.IM}
}

@ARTICLE{Scargle1982,
       author = {{Scargle}, J.~D.},
        title = "{Studies in astronomical time series analysis. II. Statistical aspects of spectral analysis of unevenly spaced data.}",
      journal = {\apj},
         year = 1982,
        month = dec,
       volume = {263},
        pages = {835-853},
          doi = {10.1086/160554},
       adsurl = {https://ui.adsabs.harvard.edu/abs/1982ApJ...263..835S}
}

@ARTICLE{Mathieu2025,
       author = {{Mathieu}, Robert D. and {Pols}, Onno R.},
        title = "{Blue Stragglers and Friends: Initial Evolutionary Pathways in Close Low-Mass Binaries}",
      journal = {\araa},
         year = 2025,
        month = aug,
       volume = {63},
       number = {1},
        pages = {467-512},
          doi = {10.1146/annurev-astro-071221-054402},
archivePrefix = {arXiv},
       eprint = {2509.20531},
 primaryClass = {astro-ph.SR},
       adsurl = {https://ui.adsabs.harvard.edu/abs/2025ARA&A..63..467M}
}

@ARTICLE{Zechmeister2009,
       author = {{Zechmeister}, M. and {K{\"u}rster}, M.},
        title = "{The generalised Lomb-Scargle periodogram. A new formalism for the floating-mean and Keplerian periodograms}",
      journal = {\aap},
         year = 2009,
        month = mar,
       volume = {496},
       number = {2},
        pages = {577-584},
          doi = {10.1051/0004-6361:200811296},
archivePrefix = {arXiv},
       eprint = {0901.2573},
 primaryClass = {astro-ph.IM},
       adsurl = {https://ui.adsabs.harvard.edu/abs/2009A&A...496..577Z}
}

@ARTICLE{astroquery,
       author = {{Ginsburg}, Adam and {Sip{\H{o}}cz}, Brigitta M. and {Brasseur}, C.~E. and {Cowperthwaite}, Philip S. and {Craig}, Matthew W. and {Deil}, Christoph and {Guillochon}, James and {Guzman}, Giannina and {Liedtke}, Simon and {Lian Lim}, Pey and {Lockhart}, Kelly E. and {Mommert}, Michael and {Morris}, Brett M. and {Norman}, Henrik and {Parikh}, Madhura and {Persson}, Magnus V. and {Robitaille}, Thomas P. and {Segovia}, Juan-Carlos and {Singer}, Leo P. and {Tollerud}, Erik J. and {de Val-Borro}, Miguel and {Valtchanov}, Ivan and {Woillez}, Julien and {Astroquery Collaboration} and {a subset of astropy Collaboration}},
        title = "{astroquery: An Astronomical Web-querying Package in Python}",
      journal = {\aj},
         year = 2019,
        month = mar,
       volume = {157},
       number = {3},
          eid = {98},
        pages = {98},
          doi = {10.3847/1538-3881/aafc33},
archivePrefix = {arXiv},
       eprint = {1901.04520},
 primaryClass = {astro-ph.IM},
       adsurl = {https://ui.adsabs.harvard.edu/abs/2019AJ....157...98G}
}

@ARTICLE{GaiaDR3,
       author = {{Gaia Collaboration} and {Brown}, A.~G.~A. and {Vallenari}, A. and {Prusti}, T. and {de Bruijne}, J.~H.~J. and {Babusiaux}, C. and {Biermann}, M. and {Creevey}, O.~L. and {Evans}, D.~W. and {Eyer}, L. and {Hutton}, A. and {Jansen}, F. and {Jordi}, C. and {Klioner}, S.~A. and {Lammers}, U. and {Lindegren}, L. and {Luri}, X. and {Mignard}, F. and {Panem}, C. and {Pourbaix}, D. and {Randich}, S. and {Sartoretti}, P. and {Soubiran}, C. and {Walton}, N.~A. and {Arenou}, F. and {Bailer-Jones}, C.~A.~L. and {Bastian}, U. and {Cropper}, M. and {Drimmel}, R. and {Katz}, D. and {Lattanzi}, M.~G. and {van Leeuwen}, F. and {Bakker}, J. and {Cacciari}, C. and {Casta{\~n}eda}, J. and {De Angeli}, F. and {Ducourant}, C. and {Fabricius}, C. and {Fouesneau}, M. and {Fr{\'e}mat}, Y. and {Guerra}, R. and {Guerrier}, A. and {Guiraud}, J. and {Jean-Antoine Piccolo}, A. and {Masana}, E. and {Messineo}, R. and {Mowlavi}, N. and {Nicolas}, C. and {Nienartowicz}, K. and {Pailler}, F. and {Panuzzo}, P. and {Riclet}, F. and {Roux}, W. and {Seabroke}, G.~M. and {Sordo}, R. and {Tanga}, P. and {Th{\'e}venin}, F. and {Gracia-Abril}, G. and {Portell}, J. and {Teyssier}, D. and {Altmann}, M. and {Andrae}, R. and {Bellas-Velidis}, I. and {Benson}, K. and {Berthier}, J. and {Blomme}, R. and {Brugaletta}, E. and {Burgess}, P.~W. and {Busso}, G. and {Carry}, B. and {Cellino}, A. and {Cheek}, N. and {Clementini}, G. and {Damerdji}, Y. and {Davidson}, M. and {Delchambre}, L. and {Dell'Oro}, A. and {Fern{\'a}ndez-Hern{\'a}ndez}, J. and {Galluccio}, L. and {Garc{\'\i}a-Lario}, P. and {Garcia-Reinaldos}, M. and {Gonz{\'a}lez-N{\'u}{\~n}ez}, J. and {Gosset}, E. and {Haigron}, R. and {Halbwachs}, J. -L. and {Hambly}, N.~C. and {Harrison}, D.~L. and {Hatzidimitriou}, D. and {Heiter}, U. and {Hern{\'a}ndez}, J. and {Hestroffer}, D. and {Hodgkin}, S.~T. and {Holl}, B. and {Jan{\ss}en}, K. and {Jevardat de Fombelle}, G. and {Jordan}, S. and {Krone-Martins}, A. and {Lanzafame}, A.~C. and {L{\"o}ffler}, W. and {Lorca}, A. and {Manteiga}, M. and {Marchal}, O. and {Marrese}, P.~M. and {Moitinho}, A. and {Mora}, A. and {Muinonen}, K. and {Osborne}, P. and {Pancino}, E. and {Pauwels}, T. and {Petit}, J. -M. and {Recio-Blanco}, A. and {Richards}, P.~J. and {Riello}, M. and {Rimoldini}, L. and {Robin}, A.~C. and {Roegiers}, T. and {Rybizki}, J. and {Sarro}, L.~M. and {Siopis}, C. and {Smith}, M. and {Sozzetti}, A. and {Ulla}, A. and {Utrilla}, E. and {van Leeuwen}, M. and {van Reeven}, W. and {Abbas}, U. and {Abreu Aramburu}, A. and {Accart}, S. and {Aerts}, C. and {Aguado}, J.~J. and {Ajaj}, M. and {Altavilla}, G. and {{\'A}lvarez}, M.~A. and {{\'A}lvarez Cid-Fuentes}, J. and {Alves}, J. and {Anderson}, R.~I. and {Anglada Varela}, E. and {Antoja}, T. and {Audard}, M. and {Baines}, D. and {Baker}, S.~G. and {Balaguer-N{\'u}{\~n}ez}, L. and {Balbinot}, E. and {Balog}, Z. and {Barache}, C. and {Barbato}, D. and {Barros}, M. and {Barstow}, M.~A. and {Bartolom{\'e}}, S. and {Bassilana}, J. -L. and {Bauchet}, N. and {Baudesson-Stella}, A. and {Becciani}, U. and {Bellazzini}, M. and {Bernet}, M. and {Bertone}, S. and {Bianchi}, L. and {Blanco-Cuaresma}, S. and {Boch}, T. and {Bombrun}, A. and {Bossini}, D. and {Bouquillon}, S. and {Bragaglia}, A. and {Bramante}, L. and {Breedt}, E. and {Bressan}, A. and {Brouillet}, N. and {Bucciarelli}, B. and {Burlacu}, A. and {Busonero}, D. and {Butkevich}, A.~G. and {Buzzi}, R. and {Caffau}, E. and {Cancelliere}, R. and {C{\'a}novas}, H. and {Cantat-Gaudin}, T. and {Carballo}, R. and {Carlucci}, T. and {Carnerero}, M.~I. and {Carrasco}, J.~M. and {Casamiquela}, L. and {Castellani}, M. and {Castro-Ginard}, A. and {Castro Sampol}, P. and {Chaoul}, L. and {Charlot}, P. and {Chemin}, L. and {Chiavassa}, A. and {Cioni}, M. -R.~L. and {Comoretto}, G. and {Cooper}, W.~J. and {Cornez}, T. and {Cowell}, S. and {Crifo}, F. and {Crosta}, M. and {Crowley}, C. and {Dafonte}, C. and {Dapergolas}, A. and {David}, M. and {David}, P. and {de Laverny}, P. and {De Luise}, F. and {De March}, R. and {De Ridder}, J. and {de Souza}, R. and {de Teodoro}, P. and {de Torres}, A. and {del Peloso}, E.~F. and {del Pozo}, E. and {Delbo}, M. and {Delgado}, A. and {Delgado}, H.~E. and {Delisle}, J. -B. and {Di Matteo}, P. and {Diakite}, S. and {Diener}, C. and {Distefano}, E. and {Dolding}, C. and {Eappachen}, D. and {Edvardsson}, B. and {Enke}, H. and {Esquej}, P. and {Fabre}, C. and {Fabrizio}, M. and {Faigler}, S. and {Fedorets}, G. and {Fernique}, P. and {Fienga}, A. and {Figueras}, F. and {Fouron}, C. and {Fragkoudi}, F. and {Fraile}, E. and {Franke}, F. and {Gai}, M. and {Garabato}, D. and {Garcia-Gutierrez}, A. and {Garc{\'\i}a-Torres}, M. and {Garofalo}, A. and {Gavras}, P. and {Gerlach}, E. and {Geyer}, R. and {Giacobbe}, P. and {Gilmore}, G. and {Girona}, S. and {Giuffrida}, G. and {Gomel}, R. and {Gomez}, A. and {Gonzalez-Santamaria}, I. and {Gonz{\'a}lez-Vidal}, J.~J. and {Granvik}, M. and {Guti{\'e}rrez-S{\'a}nchez}, R. and {Guy}, L.~P. and {Hauser}, M. and {Haywood}, M. and {Helmi}, A. and {Hidalgo}, S.~L. and {Hilger}, T. and {H{\l}adczuk}, N. and {Hobbs}, D. and {Holland}, G. and {Huckle}, H.~E. and {Jasniewicz}, G. and {Jonker}, P.~G. and {Juaristi Campillo}, J. and {Julbe}, F. and {Karbevska}, L. and {Kervella}, P. and {Khanna}, S. and {Kochoska}, A. and {Kontizas}, M. and {Kordopatis}, G. and {Korn}, A.~J. and {Kostrzewa-Rutkowska}, Z. and {Kruszy{\'n}ska}, K. and {Lambert}, S. and {Lanza}, A.~F. and {Lasne}, Y. and {Le Campion}, J. -F. and {Le Fustec}, Y. and {Lebreton}, Y. and {Lebzelter}, T. and {Leccia}, S. and {Leclerc}, N. and {Lecoeur-Taibi}, I. and {Liao}, S. and {Licata}, E. and {Lindstr{\o}m}, E.~P. and {Lister}, T.~A. and {Livanou}, E. and {Lobel}, A. and {Madrero Pardo}, P. and {Managau}, S. and {Mann}, R.~G. and {Marchant}, J.~M. and {Marconi}, M. and {Marcos Santos}, M.~M.~S. and {Marinoni}, S. and {Marocco}, F. and {Marshall}, D.~J. and {Martin Polo}, L. and {Mart{\'\i}n-Fleitas}, J.~M. and {Masip}, A. and {Massari}, D. and {Mastrobuono-Battisti}, A. and {Mazeh}, T. and {McMillan}, P.~J. and {Messina}, S. and {Michalik}, D. and {Millar}, N.~R. and {Mints}, A. and {Molina}, D. and {Molinaro}, R. and {Moln{\'a}r}, L. and {Montegriffo}, P. and {Mor}, R. and {Morbidelli}, R. and {Morel}, T. and {Morris}, D. and {Mulone}, A.~F. and {Munoz}, D. and {Muraveva}, T. and {Murphy}, C.~P. and {Musella}, I. and {Noval}, L. and {Ord{\'e}novic}, C. and {Orr{\`u}}, G. and {Osinde}, J. and {Pagani}, C. and {Pagano}, I. and {Palaversa}, L. and {Palicio}, P.~A. and {Panahi}, A. and {Pawlak}, M. and {Pe{\~n}alosa Esteller}, X. and {Penttil{\"a}}, A. and {Piersimoni}, A.~M. and {Pineau}, F. -X. and {Plachy}, E. and {Plum}, G. and {Poggio}, E. and {Poretti}, E. and {Poujoulet}, E. and {Pr{\v{s}}a}, A. and {Pulone}, L. and {Racero}, E. and {Ragaini}, S. and {Rainer}, M. and {Raiteri}, C.~M. and {Rambaux}, N. and {Ramos}, P. and {Ramos-Lerate}, M. and {Re Fiorentin}, P. and {Regibo}, S. and {Reyl{\'e}}, C. and {Ripepi}, V. and {Riva}, A. and {Rixon}, G. and {Robichon}, N. and {Robin}, C. and {Roelens}, M. and {Rohrbasser}, L. and {Romero-G{\'o}mez}, M. and {Rowell}, N. and {Royer}, F. and {Rybicki}, K.~A. and {Sadowski}, G. and {Sagrist{\`a} Sell{\'e}s}, A. and {Sahlmann}, J. and {Salgado}, J. and {Salguero}, E. and {Samaras}, N. and {Sanchez Gimenez}, V. and {Sanna}, N. and {Santove{\~n}a}, R. and {Sarasso}, M. and {Schultheis}, M. and {Sciacca}, E. and {Segol}, M. and {Segovia}, J.~C. and {S{\'e}gransan}, D. and {Semeux}, D. and {Shahaf}, S. and {Siddiqui}, H.~I. and {Siebert}, A. and {Siltala}, L. and {Slezak}, E. and {Smart}, R.~L. and {Solano}, E. and {Solitro}, F. and {Souami}, D. and {Souchay}, J. and {Spagna}, A. and {Spoto}, F. and {Steele}, I.~A. and {Steidelm{\"u}ller}, H. and {Stephenson}, C.~A. and {S{\"u}veges}, M. and {Szabados}, L. and {Szegedi-Elek}, E. and {Taris}, F. and {Tauran}, G. and {Taylor}, M.~B. and {Teixeira}, R. and {Thuillot}, W. and {Tonello}, N. and {Torra}, F. and {Torra}, J. and {Turon}, C. and {Unger}, N. and {Vaillant}, M. and {van Dillen}, E. and {Vanel}, O. and {Vecchiato}, A. and {Viala}, Y. and {Vicente}, D. and {Voutsinas}, S. and {Weiler}, M. and {Wevers}, T. and {Wyrzykowski}, {\L}. and {Yoldas}, A. and {Yvard}, P. and {Zhao}, H. and {Zorec}, J. and {Zucker}, S. and {Zurbach}, C. and {Zwitter}, T.},
        title = "{Gaia Early Data Release 3. Summary of the contents and survey properties}",
      journal = {\aap},
         year = 2021,
        month = may,
       volume = {649},
          eid = {A1},
        pages = {A1},
          doi = {10.1051/0004-6361/202039657},
archivePrefix = {arXiv},
       eprint = {2012.01533},
 primaryClass = {astro-ph.GA},
       adsurl = {https://ui.adsabs.harvard.edu/abs/2021A&A...649A...1G}
}

@INPROCEEDINGS{MAST,
       author = {{Marston}, A. and {Hargis}, J. and {Levay}, K. and {Forshay}, P. and {Mullally}, S. and {Shaw}, R.},
        title = "{Overview of the Mikulski Archive for space telescopes for the James Webb Space Telescope data archiving}",
    booktitle = {Observatory Operations: Strategies, Processes, and Systems VII},
         year = 2018,
       series = {Society of Photo-Optical Instrumentation Engineers (SPIE) Conference Series},
       volume = {10704},
        month = jul,
          eid = {1070413},
        pages = {1070413},
          doi = {10.1117/12.2311973},
       adsurl = {https://ui.adsabs.harvard.edu/abs/2018SPIE10704E..13M}
}

@article{Borucki2010,
   author = {William J. Borucki and David Koch and Gibor Basri and Natalie Batalha and Timothy Brown and Douglas Caldwell and John Caldwell and Joørgen Christensen-Dalsgaard and William D. Cochran and Edna Devore and Edward W. Dunham and Andrea K. Dupree and Thomas N. Gautier and John C. Geary and Ronald Gilliland and Alan Gould and Steve B. Howell and Jon M. Jenkins and Yoji Kondo and David W. Latham and Geoffrey W. Marcy and Soren Meibom and Hans Kjeldsen and Jack J. Lissauer and David G. Monet and David Morrison and Dimitar Sasselov and Jill Tarter and Alan Boss and Don Brownlee and Toby Owen and Derek Buzasi and David Charbonneau and Laurance Doyle and Jonathan Fortney and Eric B. Ford and Matthew J. Holman and Sara Seager and Jason H. Steffen and William F. Welsh and Jason Rowe and Howard Anderson and Lars Buchhave and David Ciardi and Lucianne Walkowicz and William Sherry and Elliott Horch and Howard Isaacson and Mark E. Everett and Debra Fischer and Guillermo Torres and John Asher Johnson and Michael Endl and Phillip MacQueen and Stephen T. Bryson and Jessie Dotson and Michael Haas and Jeffrey Kolodziejczak and Jeffrey Van Cleve and Hema Chandrasekaran and Joseph D. Twicken and Elisa V. Quintana and Bruce D. Clarke and Christopher Allen and Jie Li and Haley Wu and Peter Tenenbaum and Ekaterina Verner and Frederick Bruhweiler and Jason Barnes and Andrej Prsa},
   doi = {10.1126/science.1185402},
   issn = {00368075},
   issue = {5968},
   journal = {Science},
   title = {Kepler planet-detection mission: Introduction and first results},
   volume = {327},
   year = {2010},
}

@article{Stellingwerf1978,
   author = {R. F. Stellingwerf},
   doi = {10.1086/156444},
   issn = {0004-637X},
   journal = {The Astrophysical Journal},
   title = {Period determination using phase dispersion minimization},
   volume = {224},
   year = {1978},
}

@ARTICLE{Brewer2016,
       author = {{Brewer}, Lauren N. and {Sandquist}, Eric L. and {Mathieu}, Robert D. and {Milliman}, Katelyn and {Geller}, Aaron M. and {Jeffries}, Jr., Mark W. and {Orosz}, Jerome A. and {Brogaard}, Karsten and {Platais}, Imants and {Bruntt}, Hans and {Grundahl}, Frank and {Stello}, Dennis and {Frandsen}, S{\o}ren},
        title = "{Determining the Age of the Kepler Open Cluster NGC 6819 With a New Triple System and Other Eclipsing Binary Stars}",
      journal = {\aj},
         year = 2016,
        month = mar,
       volume = {151},
       number = {3},
          eid = {66},
        pages = {66},
          doi = {10.3847/0004-6256/151/3/66},
archivePrefix = {arXiv},
       eprint = {1601.04069},
 primaryClass = {astro-ph.SR},
       adsurl = {https://ui.adsabs.harvard.edu/abs/2016AJ....151...66B}
}

@ARTICLE{2017ApJ...837...20P,
       author = {{Price-Whelan}, Adrian M. and {Hogg}, David W. and {Foreman-Mackey}, Daniel and {Rix}, Hans-Walter},
        title = "{The Joker: A Custom Monte Carlo Sampler for Binary-star and Exoplanet Radial Velocity Data}",
      journal = {\apj},
         year = 2017,
        month = mar,
       volume = {837},
       number = {1},
          eid = {20},
        pages = {20},
          doi = {10.3847/1538-4357/aa5e50},
archivePrefix = {arXiv},
       eprint = {1610.07602},
 primaryClass = {astro-ph.SR},
       adsurl = {https://ui.adsabs.harvard.edu/abs/2017ApJ...837...20P}
}

@ARTICLE{Brown2011,
       author = {{Brown}, Timothy M. and {Latham}, David W. and {Everett}, Mark E. and {Esquerdo}, Gilbert A.},
        title = "{Kepler Input Catalog: Photometric Calibration and Stellar Classification}",
      journal = {\aj},
         year = 2011,
        month = oct,
       volume = {142},
       number = {4},
          eid = {112},
        pages = {112},
          doi = {10.1088/0004-6256/142/4/112},
archivePrefix = {arXiv},
       eprint = {1102.0342},
 primaryClass = {astro-ph.SR},
       adsurl = {https://ui.adsabs.harvard.edu/abs/2011AJ....142..112B}
}

@ARTICLE{Deliyannis2019,
       author = {{Deliyannis}, Constantine P. and {Anthony-Twarog}, Barbara J. and {Lee-Brown}, Donald B. and {Twarog}, Bruce A.},
        title = "{Li Evolution and the Open Cluster NGC 6819: A Correlation between Li Depletion and Spindown in Dwarfs More Massive Than the F-Dwarf Li-Dip}",
      journal = {\aj},
         year = 2019,
        month = oct,
       volume = {158},
       number = {4},
          eid = {163},
        pages = {163},
          doi = {10.3847/1538-3881/ab3fad},
archivePrefix = {arXiv},
       eprint = {1909.04523},
 primaryClass = {astro-ph.SR},
       adsurl = {https://ui.adsabs.harvard.edu/abs/2019AJ....158..163D}
}

@ARTICLE{Haas2010,
       author = {{Haas}, Michael R. and {Batalha}, Natalie M. and {Bryson}, Steve T. and {Caldwell}, Douglas A. and {Dotson}, Jessie L. and {Hall}, Jennifer and {Jenkins}, Jon M. and {Klaus}, Todd C. and {Koch}, David G. and {Kolodziejczak}, Jeffrey and {Middour}, Chris and {Smith}, Marcie and {Sobeck}, Charles K. and {Stober}, Jeremy and {Thompson}, Richard S. and {Van Cleve}, Jeffrey E.},
        title = "{Kepler Science Operations}",
      journal = {\apjl},
         year = 2010,
        month = apr,
       volume = {713},
       number = {2},
        pages = {L115-L119},
          doi = {10.1088/2041-8205/713/2/L115},
archivePrefix = {arXiv},
       eprint = {1001.0437},
 primaryClass = {astro-ph.EP},
       adsurl = {https://ui.adsabs.harvard.edu/abs/2010ApJ...713L.115H}
}

@ARTICLE{Alard1998,
   author = {{Alard}, C. and {Lupton}, R.~H.},
    title = "{A Method for Optimal Image Subtraction}",
  journal = {\apj},
   eprint = {astro-ph/9712287},
     year = 1998,
    month = aug,
   volume = 503,
    pages = {325-331},
      doi = {10.1086/305984},
   adsurl = {http://adsabs.harvard.edu/abs/1998ApJ...503..325A}
}

@ARTICLE{Alard2000,
   author = {{Alard}, C.},
    title = "{Image subtraction using a space-varying kernel}",
  journal = {\aaps},
     year = 2000,
    month = jun,
   volume = 144,
    pages = {363-370},
      doi = {10.1051/aas:2000214},
   adsurl = {http://adsabs.harvard.edu/abs/2000A%26AS..144..363A}
}

@article{astropy:2013,
Adsurl = {http://adsabs.harvard.edu/abs/2013A%26A...558A..33A},
Archiveprefix = {arXiv},
Author = {{Astropy Collaboration} and {Robitaille}, T.~P. and {Tollerud}, E.~J. and {Greenfield}, P. and {Droettboom}, M. and {Bray}, E. and {Aldcroft}, T. and {Davis}, M. and {Ginsburg}, A. and {Price-Whelan}, A.~M. and {Kerzendorf}, W.~E. and {Conley}, A. and {Crighton}, N. and {Barbary}, K. and {Muna}, D. and {Ferguson}, H. and {Grollier}, F. and {Parikh}, M.~M. and {Nair}, P.~H. and {Unther}, H.~M. and {Deil}, C. and {Woillez}, J. and {Conseil}, S. and {Kramer}, R. and {Turner}, J.~E.~H. and {Singer}, L. and {Fox}, R. and {Weaver}, B.~A. and {Zabalza}, V. and {Edwards}, Z.~I. and {Azalee Bostroem}, K. and {Burke}, D.~J. and {Casey}, A.~R. and {Crawford}, S.~M. and {Dencheva}, N. and {Ely}, J. and {Jenness}, T. and {Labrie}, K. and {Lim}, P.~L. and {Pierfederici}, F. and {Pontzen}, A. and {Ptak}, A. and {Refsdal}, B. and {Servillat}, M. and {Streicher}, O.},
Doi = {10.1051/0004-6361/201322068},
Eid = {A33},
Eprint = {1307.6212},
Journal = {\aap},
Month = oct,
Pages = {A33},
Primaryclass = {astro-ph.IM},
Title = {{Astropy: A community Python package for astronomy}},
Volume = 558,
Year = 2013}

@ARTICLE{astropy:2022,
       author = {{Astropy Collaboration} and {Price-Whelan}, Adrian M. and {Lim}, Pey Lian and {Earl}, Nicholas and {Starkman}, Nathaniel and {Bradley}, Larry and {Shupe}, David L. and {Patil}, Aarya A. and {Corrales}, Lia and {Brasseur}, C.~E. and {N{"o}the}, Maximilian and {Donath}, Axel and {Tollerud}, Erik and {Morris}, Brett M. and {Ginsburg}, Adam and {Vaher}, Eero and {Weaver}, Benjamin A. and {Tocknell}, James and {Jamieson}, William and {van Kerkwijk}, Marten H. and {Robitaille}, Thomas P. and {Merry}, Bruce and {Bachetti}, Matteo and {G{"u}nther}, H. Moritz and {Aldcroft}, Thomas L. and {Alvarado-Montes}, Jaime A. and {Archibald}, Anne M. and {B{'o}di}, Attila and {Bapat}, Shreyas and {Barentsen}, Geert and {Baz{'a}n}, Juanjo and {Biswas}, Manish and {Boquien}, M{'e}d{'e}ric and {Burke}, D.~J. and {Cara}, Daria and {Cara}, Mihai and {Conroy}, Kyle E. and {Conseil}, Simon and {Craig}, Matthew W. and {Cross}, Robert M. and {Cruz}, Kelle L. and {D'Eugenio}, Francesco and {Dencheva}, Nadia and {Devillepoix}, Hadrien A.~R. and {Dietrich}, J{"o}rg P. and {Eigenbrot}, Arthur Davis and {Erben}, Thomas and {Ferreira}, Leonardo and {Foreman-Mackey}, Daniel and {Fox}, Ryan and {Freij}, Nabil and {Garg}, Suyog and {Geda}, Robel and {Glattly}, Lauren and {Gondhalekar}, Yash and {Gordon}, Karl D. and {Grant}, David and {Greenfield}, Perry and {Groener}, Austen M. and {Guest}, Steve and {Gurovich}, Sebastian and {Handberg}, Rasmus and {Hart}, Akeem and {Hatfield-Dodds}, Zac and {Homeier}, Derek and {Hosseinzadeh}, Griffin and {Jenness}, Tim and {Jones}, Craig K. and {Joseph}, Prajwel and {Kalmbach}, J. Bryce and {Karamehmetoglu}, Emir and {Ka{l}uszy{'n}ski}, Miko{l}aj and {Kelley}, Michael S.~P. and {Kern}, Nicholas and {Kerzendorf}, Wolfgang E. and {Koch}, Eric W. and {Kulumani}, Shankar and {Lee}, Antony and {Ly}, Chun and {Ma}, Zhiyuan and {MacBride}, Conor and {Maljaars}, Jakob M. and {Muna}, Demitri and {Murphy}, N.~A. and {Norman}, Henrik and {O'Steen}, Richard and {Oman}, Kyle A. and {Pacifici}, Camilla and {Pascual}, Sergio and {Pascual-Granado}, J. and {Patil}, Rohit R. and {Perren}, Gabriel I. and {Pickering}, Timothy E. and {Rastogi}, Tanuj and {Roulston}, Benjamin R. and {Ryan}, Daniel F. and {Rykoff}, Eli S. and {Sabater}, Jose and {Sakurikar}, Parikshit and {Salgado}, Jes{'u}s and {Sanghi}, Aniket and {Saunders}, Nicholas and {Savchenko}, Volodymyr and {Schwardt}, Ludwig and {Seifert-Eckert}, Michael and {Shih}, Albert Y. and {Jain}, Anany Shrey and {Shukla}, Gyanendra and {Sick}, Jonathan and {Simpson}, Chris and {Singanamalla}, Sudheesh and {Singer}, Leo P. and {Singhal}, Jaladh and {Sinha}, Manodeep and {Sip{H{o}}cz}, Brigitta M. and {Spitler}, Lee R. and {Stansby}, David and {Streicher}, Ole and {{{S}}umak}, Jani and {Swinbank}, John D. and {Taranu}, Dan S. and {Tewary}, Nikita and {Tremblay}, Grant R. and {Val-Borro}, Miguel de and {Van Kooten}, Samuel J. and {Vasovi{'c}}, Zlatan and {Verma}, Shresth and {de Miranda Cardoso}, Jos{'e} Vin{'i}cius and {Williams}, Peter K.~G. and {Wilson}, Tom J. and {Winkel}, Benjamin and {Wood-Vasey}, W.~M. and {Xue}, Rui and {Yoachim}, Peter and {Zhang}, Chen and {Zonca}, Andrea and {Astropy Project Contributors}},
        title = "{The Astropy Project: Sustaining and Growing a Community-oriented Open-source Project and the Latest Major Release (v5.0) of the Core Package}",
      journal = {\apj},
         year = 2022,
        month = aug,
       volume = {935},
       number = {2},
          eid = {167},
        pages = {167},
          doi = {10.3847/1538-4357/ac7c74},
archivePrefix = {arXiv},
       eprint = {2206.14220},
 primaryClass = {astro-ph.IM},
       adsurl = {https://ui.adsabs.harvard.edu/abs/2022ApJ...935..167A}
}

@ARTICLE{astropy2018,
       author = {{Astropy Collaboration} and {Price-Whelan}, A.~M. and
         {Sip{\H{o}}cz}, B.~M. and {G{\"u}nther}, H.~M. and {Lim}, P.~L. and
         {Crawford}, S.~M. and {Conseil}, S. and {Shupe}, D.~L. and
         {Craig}, M.~W. and {Dencheva}, N.},
        title = "{The Astropy Project: Building an Open-science Project and Status of the v2.0 Core Package}",
      journal = {\aj},
         year = "2018",
        month = "Sep",
       volume = {156},
       number = {3},
          eid = {123},
        pages = {123},
          doi = {10.3847/1538-3881/aabc4f},
archivePrefix = {arXiv},
       eprint = {1801.02634},
 primaryClass = {astro-ph.IM},
       adsurl = {https://ui.adsabs.harvard.edu/abs/2018AJ....156..123A}
}

@MISC{Bhatti2018,
       author = {{Bhatti}, W. and {Bouma}, L. and {Wallace}, J. and {Price-Whelan}, A.},
        title = "{Waqasbhatti/Astrobase: Astrobase V0.3.14}",
         year = "2018",
        month = "May",
          eid = {10.5281/zenodo.1246769},
          doi = {10.5281/zenodo.1246769},
      version = {v0.3.14},
    publisher = {Zenodo},
       adsurl = {https://ui.adsabs.harvard.edu/abs/2018zndo...1246769B}
}

@ARTICLE{CantatGaudin2020,
       author = {{Cantat-Gaudin}, T. and {Anders}, F.},
        title = "{Clusters and mirages: cataloguing stellar aggregates in the Milky Way}",
      journal = {\aap},
         year = 2020,
        month = jan,
       volume = {633},
          eid = {A99},
        pages = {A99},
          doi = {10.1051/0004-6361/201936691},
archivePrefix = {arXiv},
       eprint = {1911.07075},
 primaryClass = {astro-ph.SR},
       adsurl = {https://ui.adsabs.harvard.edu/abs/2020A&A...633A..99C}
}

@ARTICLE{Cantat2018,
   author = {{Cantat-Gaudin}, T. and {Jordi}, C. and {Vallenari}, A. and 
	{Bragaglia}, A. and {Balaguer-N{\'u}{\~n}ez}, L. and {Soubiran}, C. and 
	{Bossini}, D. and {Moitinho}, A. and {Castro-Ginard}, A. and 
	{Krone-Martins}, A. and {Casamiquela}, L. and {Sordo}, R. and 
	{Carrera}, R.},
    title = "{A Gaia DR2 view of the Open Cluster population in the Milky Way}",
  journal = {ArXiv e-prints},
archivePrefix = "arXiv",
   eprint = {1805.08726},
     year = 2018,
    month = may,
   adsurl = {http://adsabs.harvard.edu/abs/2018arXiv180508726C}
}

@ARTICLE{Huang2015,
   author = {{Huang}, C.~X. and {Penev}, K. and {Hartman}, J.~D. and {Bakos}, G.~{\'A}. and 
	{Bhatti}, W. and {Domsa}, I. and {de Val-Borro}, M.},
    title = "{High-precision photometry for K2 Campaign 1}",
  journal = {\mnras},
archivePrefix = "arXiv",
   eprint = {1507.07578},
 primaryClass = "astro-ph.EP",
     year = 2015,
    month = dec,
   volume = 454,
    pages = {4159-4171},
      doi = {10.1093/mnras/stv2257},
   adsurl = {http://adsabs.harvard.edu/abs/2015MNRAS.454.4159H}
}

@BOOK{Karttunen2017,
       author = {{Karttunen}, Hannu and {Kr{\"o}ger}, Pekka and {Oja}, Heikki and {Poutanen}, Markku and {Donner}, Karl Johan},
        title = "{Fundamental Astronomy}",
         year = 2017,
          doi = {10.1007/978-3-662-53045-0},
       adsurl = {https://ui.adsabs.harvard.edu/abs/2017fuas.book.....K}
}

@ARTICLE{Kovacs2005,
       author = {{Kov{\'a}cs}, G{\'e}za and {Bakos}, G{\'a}sp{\'a}r and {Noyes}, Robert
        W.},
        title = "{A trend filtering algorithm for wide-field variability surveys}",
      journal = {\mnras},
         year = 2005,
        month = Jan,
       volume = {356},
        pages = {557-567},
          doi = {10.1111/j.1365-2966.2004.08479.x},
archivePrefix = {arXiv},
       eprint = {astro-ph/0411724},
 primaryClass = {astro-ph},
       adsurl = {https://ui.adsabs.harvard.edu/#abs/2005MNRAS.356..557K}
}

@ARTICLE{Fabricius2021,
       author = {{Fabricius}, C. and {Luri}, X. and {Arenou}, F. and {Babusiaux}, C. and {Helmi}, A. and {Muraveva}, T. and {Reyl{\'e}}, C. and {Spoto}, F. and {Vallenari}, A. and {Antoja}, T. and {Balbinot}, E. and {Barache}, C. and {Bauchet}, N. and {Bragaglia}, A. and {Busonero}, D. and {Cantat-Gaudin}, T. and {Carrasco}, J.~M. and {Diakit{\'e}}, S. and {Fabrizio}, M. and {Figueras}, F. and {Garcia-Gutierrez}, A. and {Garofalo}, A. and {Jordi}, C. and {Kervella}, P. and {Khanna}, S. and {Leclerc}, N. and {Licata}, E. and {Lambert}, S. and {Marrese}, P.~M. and {Masip}, A. and {Ramos}, P. and {Robichon}, N. and {Robin}, A.~C. and {Romero-G{\'o}mez}, M. and {Rubele}, S. and {Weiler}, M.},
        title = "{Gaia Early Data Release 3. Catalogue validation}",
      journal = {\aap},
         year = 2021,
        month = may,
       volume = {649},
          eid = {A5},
        pages = {A5},
          doi = {10.1051/0004-6361/202039834},
archivePrefix = {arXiv},
       eprint = {2012.06242},
 primaryClass = {astro-ph.GA},
       adsurl = {https://ui.adsabs.harvard.edu/abs/2021A&A...649A...5F}
}

@ARTICLE{2020SciPy-NMeth,
  author  = {Virtanen, Pauli and Gommers, Ralf and Oliphant, Travis E. and
            Haberland, Matt and Reddy, Tyler and Cournapeau, David and
            Burovski, Evgeni and Peterson, Pearu and Weckesser, Warren and
            Bright, Jonathan and {van der Walt}, St{\'e}fan J. and
            Brett, Matthew and Wilson, Joshua and Millman, K. Jarrod and
            Mayorov, Nikolay and Nelson, Andrew R. J. and Jones, Eric and
            Kern, Robert and Larson, Eric and Carey, C J and
            Polat, {\.I}lhan and Feng, Yu and Moore, Eric W. and
            {VanderPlas}, Jake and Laxalde, Denis and Perktold, Josef and
            Cimrman, Robert and Henriksen, Ian and Quintero, E. A. and
            Harris, Charles R. and Archibald, Anne M. and
            Ribeiro, Ant{\^o}nio H. and Pedregosa, Fabian and
            {van Mulbregt}, Paul and {SciPy 1.0 Contributors}},
  title   = {{{SciPy} 1.0: Fundamental Algorithms for Scientific
            Computing in Python}},
  journal = {Nature Methods},
  year    = {2020},
  volume  = {17},
  pages   = {261--272},
  adsurl  = {https://rdcu.be/b08Wh},
  doi     = {10.1038/s41592-019-0686-2},
}

@ARTICLE{BailerJones2021,
       author = {{Bailer-Jones}, C.~A.~L. and {Rybizki}, J. and {Fouesneau}, M. and {Demleitner}, M. and {Andrae}, R.},
        title = "{Estimating Distances from Parallaxes. V. Geometric and Photogeometric Distances to 1.47 Billion Stars in Gaia Early Data Release 3}",
      journal = {\aj},
         year = 2021,
        month = mar,
       volume = {161},
       number = {3},
          eid = {147},
        pages = {147},
          doi = {10.3847/1538-3881/abd806},
archivePrefix = {arXiv},
       eprint = {2012.05220},
 primaryClass = {astro-ph.SR},
       adsurl = {https://ui.adsabs.harvard.edu/abs/2021AJ....161..147B}
}

@ARTICLE{matplotlib,
       author = {{Hunter}, John D.},
        title = "{Matplotlib: A 2D Graphics Environment}",
      journal = {Computing in Science and Engineering},
         year = "2007",
        month = "May",
       volume = {9},
       number = {3},
        pages = {90-95},
          doi = {10.1109/MCSE.2007.55},
       adsurl = {https://ui.adsabs.harvard.edu/abs/2007CSE.....9...90H}
}

@ARTICLE{Sagynbayeva2025,
       author = {{Sagynbayeva}, Sabina and {Colman}, Isabel L. and {Farr}, Will M.},
        title = "{Rotation Periods for Stars in Open Cluster NGC 6819 From Kepler IRIS Light Curves}",
      journal = {arXiv e-prints},
         year = 2025,
        month = oct,
          eid = {arXiv:2510.02255},
        pages = {arXiv:2510.02255},
          doi = {10.48550/arXiv.2510.02255},
archivePrefix = {arXiv},
       eprint = {2510.02255},
 primaryClass = {astro-ph.SR},
       adsurl = {https://ui.adsabs.harvard.edu/abs/2025arXiv251002255S}
}

@ARTICLE{Angus2018,
       author = {{Angus}, Ruth and {Morton}, Timothy and {Aigrain}, Suzanne and {Foreman-Mackey}, Daniel and {Rajpaul}, Vinesh},
        title = "{Inferring probabilistic stellar rotation periods using Gaussian processes}",
      journal = {\mnras},
         year = 2018,
        month = feb,
       volume = {474},
       number = {2},
        pages = {2094-2108},
          doi = {10.1093/mnras/stx2109},
archivePrefix = {arXiv},
       eprint = {1706.05459},
 primaryClass = {astro-ph.SR},
       adsurl = {https://ui.adsabs.harvard.edu/abs/2018MNRAS.474.2094A}
}

@ARTICLE{2018A&A...619A.176B,
       author = {{Bragaglia}, Angela and {Fu}, Xiaoting and {Mucciarelli}, Alessio and {Andreuzzi}, Gloria and {Donati}, Paolo},
        title = "{The chemical composition of the oldest nearby open cluster Ruprecht 147}",
      journal = {\aap},
         year = 2018,
        month = nov,
       volume = {619},
          eid = {A176},
        pages = {A176},
          doi = {10.1051/0004-6361/201833888},
archivePrefix = {arXiv},
       eprint = {1809.06868},
 primaryClass = {astro-ph.SR},
       adsurl = {https://ui.adsabs.harvard.edu/abs/2018A&A...619A.176B}
}

@ARTICLE{Curtis2020,
       author = {{Curtis}, Jason Lee and {Ag{\"u}eros}, Marcel A. and {Matt}, Sean P. and {Covey}, Kevin R. and {Douglas}, Stephanie T. and {Angus}, Ruth and {Saar}, Steven H. and {Cody}, Ann Marie and {Vanderburg}, Andrew and {Law}, Nicholas M. and {Kraus}, Adam L. and {Latham}, David W. and {Baranec}, Christoph and {Riddle}, Reed and {Ziegler}, Carl and {Lund}, Mikkel N. and {Torres}, Guillermo and {Meibom}, S{\o}ren and {Aguirre}, Victor Silva and {Wright}, Jason T.},
        title = "{When Do Stalled Stars Resume Spinning Down? Advancing Gyrochronology with Ruprecht 147}",
      journal = {\apj},
         year = 2020,
        month = dec,
       volume = {904},
       number = {2},
          eid = {140},
        pages = {140},
          doi = {10.3847/1538-4357/abbf58},
archivePrefix = {arXiv},
       eprint = {2010.02272},
 primaryClass = {astro-ph.SR},
       adsurl = {https://ui.adsabs.harvard.edu/abs/2020ApJ...904..140C}
}

@misc{smplotlib,
  author       = {Jiaxuan Li},
  title        = {AstroJacobLi/smplotlib: v0.0.9},
  month        = jul,
  year         = 2023,
  publisher    = {Zenodo},
  version      = {v0.0.9},
  doi          = {10.5281/zenodo.8126529},
  url          = {https://doi.org/10.5281/zenodo.8126529},
}

@ARTICLE{Chen2020,
       author = {{Chen}, Xiaodian and {Wang}, Shu and {Deng}, Licai and {de Grijs}, Richard and {Yang}, Ming and {Tian}, Hao},
        title = "{The Zwicky Transient Facility Catalog of Periodic Variable Stars}",
      journal = {\apjs},
         year = 2020,
        month = jul,
       volume = {249},
       number = {1},
          eid = {18},
        pages = {18},
          doi = {10.3847/1538-4365/ab9cae},
archivePrefix = {arXiv},
       eprint = {2005.08662},
 primaryClass = {astro-ph.SR},
       adsurl = {https://ui.adsabs.harvard.edu/abs/2020ApJS..249...18C}
}

@ARTICLE{Reinhold2023,
       author = {{Reinhold}, Timo and {Shapiro}, Alexander I. and {Solanki}, Sami K. and {Basri}, Gibor},
        title = "{New rotation period measurements of 67 163 Kepler stars}",
      journal = {\aap},
         year = 2023,
        month = oct,
       volume = {678},
          eid = {A24},
        pages = {A24},
          doi = {10.1051/0004-6361/202346789},
archivePrefix = {arXiv},
       eprint = {2308.04272},
 primaryClass = {astro-ph.SR},
       adsurl = {https://ui.adsabs.harvard.edu/abs/2023A&A...678A..24R}
}

@ARTICLE{Thompson2015,
       author = {{Thompson}, Susan E. and {Mullally}, Fergal and {Coughlin}, Jeff and {Christiansen}, Jessie L. and {Henze}, Christopher E. and {Haas}, Michael R. and {Burke}, Christopher J.},
        title = "{A Machine Learning Technique to Identify Transit Shaped Signals}",
      journal = {\apj},
         year = 2015,
        month = oct,
       volume = {812},
       number = {1},
          eid = {46},
        pages = {46},
          doi = {10.1088/0004-637X/812/1/46},
archivePrefix = {arXiv},
       eprint = {1509.00041},
 primaryClass = {astro-ph.EP},
       adsurl = {https://ui.adsabs.harvard.edu/abs/2015ApJ...812...46T}
}

@ARTICLE{Santos2021,
       author = {{Santos}, A.~R.~G. and {Breton}, S.~N. and {Mathur}, S. and {Garc{\'\i}a}, R.~A.},
        title = "{Surface Rotation and Photometric Activity for Kepler Targets. II. G and F Main-sequence Stars and Cool Subgiant Stars}",
      journal = {\apjs},
         year = 2021,
        month = jul,
       volume = {255},
       number = {1},
          eid = {17},
        pages = {17},
          doi = {10.3847/1538-4365/ac033f},
archivePrefix = {arXiv},
       eprint = {2107.02217},
 primaryClass = {astro-ph.SR},
       adsurl = {https://ui.adsabs.harvard.edu/abs/2021ApJS..255...17S}
}

@ARTICLE{Reinhold_2015,
       author = {{Reinhold}, Timo and {Gizon}, Laurent},
        title = "{Rotation, differential rotation, and gyrochronology of active Kepler stars}",
      journal = {\aap},
         year = 2015,
        month = nov,
       volume = {583},
          eid = {A65},
        pages = {A65},
          doi = {10.1051/0004-6361/201526216},
       adsurl = {https://ui.adsabs.harvard.edu/abs/2015A&A...583A..65R},
}

@ARTICLE{Meibom2015,
       author = {{Meibom}, S{\o}ren and {Barnes}, Sydney A. and {Platais}, Imants and
        {Gilliland}, Ronald L. and {Latham}, David W. and {Mathieu},
        Robert D.},
        title = "{A spin-down clock for cool stars from observations of a 2.5-billion-
        year-old cluster}",
      journal = {\nat},
         year = 2015,
        month = Jan,
       volume = {517},
        pages = {589-591},
          doi = {10.1038/nature14118},
 primaryClass = {astro-ph.SR},
       adsurl = {https://ui.adsabs.harvard.edu/#abs/2015Natur.517..589M}
}

@ARTICLE{GaiaCollaboration2021,
       author = {{Gaia Collaboration} and {Brown}, A.~G.~A. and {Vallenari}, A. and {Prusti}, T. and {de Bruijne}, J.~H.~J. and {Babusiaux}, C. and {Biermann}, M. and {Creevey}, O.~L. and {Evans}, D.~W. and {Eyer}, L. and {Hutton}, A. and {Jansen}, F. and {Jordi}, C. and {Klioner}, S.~A. and {Lammers}, U. and {Lindegren}, L. and {Luri}, X. and {Mignard}, F. and {Panem}, C. and {Pourbaix}, D. and {Randich}, S. and {Sartoretti}, P. and {Soubiran}, C. and {Walton}, N.~A. and {Arenou}, F. and {Bailer-Jones}, C.~A.~L. and {Bastian}, U. and {Cropper}, M. and {Drimmel}, R. and {Katz}, D. and {Lattanzi}, M.~G. and {van Leeuwen}, F. and {Bakker}, J. and {Cacciari}, C. and {Casta{\~n}eda}, J. and {De Angeli}, F. and {Ducourant}, C. and {Fabricius}, C. and {Fouesneau}, M. and {Fr{\'e}mat}, Y. and {Guerra}, R. and {Guerrier}, A. and {Guiraud}, J. and {Jean-Antoine Piccolo}, A. and {Masana}, E. and {Messineo}, R. and {Mowlavi}, N. and {Nicolas}, C. and {Nienartowicz}, K. and {Pailler}, F. and {Panuzzo}, P. and {Riclet}, F. and {Roux}, W. and {Seabroke}, G.~M. and {Sordo}, R. and {Tanga}, P. and {Th{\'e}venin}, F. and {Gracia-Abril}, G. and {Portell}, J. and {Teyssier}, D. and {Altmann}, M. and {Andrae}, R. and {Bellas-Velidis}, I. and {Benson}, K. and {Berthier}, J. and {Blomme}, R. and {Brugaletta}, E. and {Burgess}, P.~W. and {Busso}, G. and {Carry}, B. and {Cellino}, A. and {Cheek}, N. and {Clementini}, G. and {Damerdji}, Y. and {Davidson}, M. and {Delchambre}, L. and {Dell'Oro}, A. and {Fern{\'a}ndez-Hern{\'a}ndez}, J. and {Galluccio}, L. and {Garc{\'\i}a-Lario}, P. and {Garcia-Reinaldos}, M. and {Gonz{\'a}lez-N{\'u}{\~n}ez}, J. and {Gosset}, E. and {Haigron}, R. and {Halbwachs}, J. -L. and {Hambly}, N.~C. and {Harrison}, D.~L. and {Hatzidimitriou}, D. and {Heiter}, U. and {Hern{\'a}ndez}, J. and {Hestroffer}, D. and {Hodgkin}, S.~T. and {Holl}, B. and {Jan{\ss}en}, K. and {Jevardat de Fombelle}, G. and {Jordan}, S. and {Krone-Martins}, A. and {Lanzafame}, A.~C. and {L{\"o}ffler}, W. and {Lorca}, A. and {Manteiga}, M. and {Marchal}, O. and {Marrese}, P.~M. and {Moitinho}, A. and {Mora}, A. and {Muinonen}, K. and {Osborne}, P. and {Pancino}, E. and {Pauwels}, T. and {Petit}, J. -M. and {Recio-Blanco}, A. and {Richards}, P.~J. and {Riello}, M. and {Rimoldini}, L. and {Robin}, A.~C. and {Roegiers}, T. and {Rybizki}, J. and {Sarro}, L.~M. and {Siopis}, C. and {Smith}, M. and {Sozzetti}, A. and {Ulla}, A. and {Utrilla}, E. and {van Leeuwen}, M. and {van Reeven}, W. and {Abbas}, U. and {Abreu Aramburu}, A. and {Accart}, S. and {Aerts}, C. and {Aguado}, J.~J. and {Ajaj}, M. and {Altavilla}, G. and {{\'A}lvarez}, M.~A. and {{\'A}lvarez Cid-Fuentes}, J. and {Alves}, J. and {Anderson}, R.~I. and {Anglada Varela}, E. and {Antoja}, T. and {Audard}, M. and {Baines}, D. and {Baker}, S.~G. and {Balaguer-N{\'u}{\~n}ez}, L. and {Balbinot}, E. and {Balog}, Z. and {Barache}, C. and {Barbato}, D. and {Barros}, M. and {Barstow}, M.~A. and {Bartolom{\'e}}, S. and {Bassilana}, J. -L. and {Bauchet}, N. and {Baudesson-Stella}, A. and {Becciani}, U. and {Bellazzini}, M. and {Bernet}, M. and {Bertone}, S. and {Bianchi}, L. and {Blanco-Cuaresma}, S. and {Boch}, T. and {Bombrun}, A. and {Bossini}, D. and {Bouquillon}, S. and {Bragaglia}, A. and {Bramante}, L. and {Breedt}, E. and {Bressan}, A. and {Brouillet}, N. and {Bucciarelli}, B. and {Burlacu}, A. and {Busonero}, D. and {Butkevich}, A.~G. and {Buzzi}, R. and {Caffau}, E. and {Cancelliere}, R. and {C{\'a}novas}, H. and {Cantat-Gaudin}, T. and {Carballo}, R. and {Carlucci}, T. and {Carnerero}, M.~I. and {Carrasco}, J.~M. and {Casamiquela}, L. and {Castellani}, M. and {Castro-Ginard}, A. and {Castro Sampol}, P. and {Chaoul}, L. and {Charlot}, P. and {Chemin}, L. and {Chiavassa}, A. and {Cioni}, M. -R.~L. and {Comoretto}, G. and {Cooper}, W.~J. and {Cornez}, T. and {Cowell}, S. and {Crifo}, F. and {Crosta}, M. and {Crowley}, C. and {Dafonte}, C. and {Dapergolas}, A. and {David}, M. and {David}, P. and {de Laverny}, P. and {De Luise}, F. and {De March}, R. and {De Ridder}, J. and {de Souza}, R. and {de Teodoro}, P. and {de Torres}, A. and {del Peloso}, E.~F. and {del Pozo}, E. and {Delbo}, M. and {Delgado}, A. and {Delgado}, H.~E. and {Delisle}, J. -B. and {Di Matteo}, P. and {Diakite}, S. and {Diener}, C. and {Distefano}, E. and {Dolding}, C. and {Eappachen}, D. and {Edvardsson}, B. and {Enke}, H. and {Esquej}, P. and {Fabre}, C. and {Fabrizio}, M. and {Faigler}, S. and {Fedorets}, G. and {Fernique}, P. and {Fienga}, A. and {Figueras}, F. and {Fouron}, C. and {Fragkoudi}, F. and {Fraile}, E. and {Franke}, F. and {Gai}, M. and {Garabato}, D. and {Garcia-Gutierrez}, A. and {Garc{\'\i}a-Torres}, M. and {Garofalo}, A. and {Gavras}, P. and {Gerlach}, E. and {Geyer}, R. and {Giacobbe}, P. and {Gilmore}, G. and {Girona}, S. and {Giuffrida}, G. and {Gomel}, R. and {Gomez}, A. and {Gonzalez-Santamaria}, I. and {Gonz{\'a}lez-Vidal}, J.~J. and {Granvik}, M. and {Guti{\'e}rrez-S{\'a}nchez}, R. and {Guy}, L.~P. and {Hauser}, M. and {Haywood}, M. and {Helmi}, A. and {Hidalgo}, S.~L. and {Hilger}, T. and {H{\l}adczuk}, N. and {Hobbs}, D. and {Holland}, G. and {Huckle}, H.~E. and {Jasniewicz}, G. and {Jonker}, P.~G. and {Juaristi Campillo}, J. and {Julbe}, F. and {Karbevska}, L. and {Kervella}, P. and {Khanna}, S. and {Kochoska}, A. and {Kontizas}, M. and {Kordopatis}, G. and {Korn}, A.~J. and {Kostrzewa-Rutkowska}, Z. and {Kruszy{\'n}ska}, K. and {Lambert}, S. and {Lanza}, A.~F. and {Lasne}, Y. and {Le Campion}, J. -F. and {Le Fustec}, Y. and {Lebreton}, Y. and {Lebzelter}, T. and {Leccia}, S. and {Leclerc}, N. and {Lecoeur-Taibi}, I. and {Liao}, S. and {Licata}, E. and {Lindstr{\o}m}, E.~P. and {Lister}, T.~A. and {Livanou}, E. and {Lobel}, A. and {Madrero Pardo}, P. and {Managau}, S. and {Mann}, R.~G. and {Marchant}, J.~M. and {Marconi}, M. and {Marcos Santos}, M.~M.~S. and {Marinoni}, S. and {Marocco}, F. and {Marshall}, D.~J. and {Martin Polo}, L. and {Mart{\'\i}n-Fleitas}, J.~M. and {Masip}, A. and {Massari}, D. and {Mastrobuono-Battisti}, A. and {Mazeh}, T. and {McMillan}, P.~J. and {Messina}, S. and {Michalik}, D. and {Millar}, N.~R. and {Mints}, A. and {Molina}, D. and {Molinaro}, R. and {Moln{\'a}r}, L. and {Montegriffo}, P. and {Mor}, R. and {Morbidelli}, R. and {Morel}, T. and {Morris}, D. and {Mulone}, A.~F. and {Munoz}, D. and {Muraveva}, T. and {Murphy}, C.~P. and {Musella}, I. and {Noval}, L. and {Ord{\'e}novic}, C. and {Orr{\`u}}, G. and {Osinde}, J. and {Pagani}, C. and {Pagano}, I. and {Palaversa}, L. and {Palicio}, P.~A. and {Panahi}, A. and {Pawlak}, M. and {Pe{\~n}alosa Esteller}, X. and {Penttil{\"a}}, A. and {Piersimoni}, A.~M. and {Pineau}, F. -X. and {Plachy}, E. and {Plum}, G. and {Poggio}, E. and {Poretti}, E. and {Poujoulet}, E. and {Pr{\v{s}}a}, A. and {Pulone}, L. and {Racero}, E. and {Ragaini}, S. and {Rainer}, M. and {Raiteri}, C.~M. and {Rambaux}, N. and {Ramos}, P. and {Ramos-Lerate}, M. and {Re Fiorentin}, P. and {Regibo}, S. and {Reyl{\'e}}, C. and {Ripepi}, V. and {Riva}, A. and {Rixon}, G. and {Robichon}, N. and {Robin}, C. and {Roelens}, M. and {Rohrbasser}, L. and {Romero-G{\'o}mez}, M. and {Rowell}, N. and {Royer}, F. and {Rybicki}, K.~A. and {Sadowski}, G. and {Sagrist{\`a} Sell{\'e}s}, A. and {Sahlmann}, J. and {Salgado}, J. and {Salguero}, E. and {Samaras}, N. and {Sanchez Gimenez}, V. and {Sanna}, N. and {Santove{\~n}a}, R. and {Sarasso}, M. and {Schultheis}, M. and {Sciacca}, E. and {Segol}, M. and {Segovia}, J.~C. and {S{\'e}gransan}, D. and {Semeux}, D. and {Shahaf}, S. and {Siddiqui}, H.~I. and {Siebert}, A. and {Siltala}, L. and {Slezak}, E. and {Smart}, R.~L. and {Solano}, E. and {Solitro}, F. and {Souami}, D. and {Souchay}, J. and {Spagna}, A. and {Spoto}, F. and {Steele}, I.~A. and {Steidelm{\"u}ller}, H. and {Stephenson}, C.~A. and {S{\"u}veges}, M. and {Szabados}, L. and {Szegedi-Elek}, E. and {Taris}, F. and {Tauran}, G. and {Taylor}, M.~B. and {Teixeira}, R. and {Thuillot}, W. and {Tonello}, N. and {Torra}, F. and {Torra}, J. and {Turon}, C. and {Unger}, N. and {Vaillant}, M. and {van Dillen}, E. and {Vanel}, O. and {Vecchiato}, A. and {Viala}, Y. and {Vicente}, D. and {Voutsinas}, S. and {Weiler}, M. and {Wevers}, T. and {Wyrzykowski}, {\L}. and {Yoldas}, A. and {Yvard}, P. and {Zhao}, H. and {Zorec}, J. and {Zucker}, S. and {Zurbach}, C. and {Zwitter}, T.},
        title = "{Gaia Early Data Release 3. Summary of the contents and survey properties}",
      journal = {\aap},
         year = 2021,
        month = may,
       volume = {649},
          eid = {A1},
        pages = {A1},
          doi = {10.1051/0004-6361/202039657},
archivePrefix = {arXiv},
       eprint = {2012.01533},
 primaryClass = {astro-ph.GA},
       adsurl = {https://ui.adsabs.harvard.edu/abs/2021A&A...649A...1G}
}

@misc{Jayasinghe2020,
       author = {{Jayasinghe}, T. and {Kochanek}, C.~S. and {Stanek}, K.~Z. and {Shappee}, B.~J. and {Holoien}, T.~W.-S. and {Thompson}, T.~A. and {Prieto}, J.~L. and {Dong}, S. and {Pawlak}, M. and {Shields}, J.~V. and {Pojmanski}, G. and {Otero}, S. and {Britt}, C.~A. and {Will}, D.},
        title = "{VizieR Online Data Catalog: ASAS-SN catalog of variable stars (Jayasinghe+, 2018-2020)}",
 howpublished = {VizieR On-line Data Catalog: II/366.  Originally published in: 2018MNRAS.477.3145J},
         year = 2020,
        month = oct,
          eid = {II/366},
       adsurl = {https://ui.adsabs.harvard.edu/abs/2020yCat.2366....0J}
}

@ARTICLE{Lurie2017,
       author = {{Lurie}, John C. and {Vyhmeister}, Karl and {Hawley}, Suzanne L. and {Adilia}, Jamel and {Chen}, Andrea and {Davenport}, James R.~A. and {Juri{\'c}}, Mario and {Puig-Holzman}, Michael and {Weisenburger}, Kolby L.},
        title = "{Tidal Synchronization and Differential Rotation of Kepler Eclipsing Binaries}",
      journal = {\aj},
         year = 2017,
        month = dec,
       volume = {154},
       number = {6},
          eid = {250},
        pages = {250},
          doi = {10.3847/1538-3881/aa974d},
archivePrefix = {arXiv},
       eprint = {1710.07339},
 primaryClass = {astro-ph.SR},
       adsurl = {https://ui.adsabs.harvard.edu/abs/2017AJ....154..250L}
}

@ARTICLE{Gavras2023,
       author = {{Gavras}, Panagiotis and {Rimoldini}, Lorenzo and {Nienartowicz}, Krzysztof and {de Fombelle}, Gr{\'e}gory Jevardat and {Holl}, Berry and {{\'A}brah{\'a}m}, P{\'e}ter and {Audard}, Marc and {Carnerero}, Maria I. and {Clementini}, Gisella and {De Ridder}, Joris and {Distefano}, Elisa and {Garcia-Lario}, Pedro and {Garofalo}, Alessia and {K{\'o}sp{\'a}l}, {\'A}gnes and {Kruszy{\'n}ska}, Katarzyna and {Kun}, M{\'a}ria and {Lecoeur-Ta{\"\i}bi}, Isabelle and {Marton}, G{\'a}bor and {Mazeh}, Tsevi and {Mowlavi}, Nami and {Raiteri}, Claudia M. and {Ripepi}, Vincenzo and {Szabados}, L{\'a}szl{\'o} and {Zucker}, Shay and {Eyer}, Laurent},
        title = "{Gaia Data Release 3. Cross-match of Gaia sources with variable objects from the literature}",
      journal = {\aap},
         year = 2023,
        month = jun,
       volume = {674},
          eid = {A22},
        pages = {A22},
          doi = {10.1051/0004-6361/202244367},
archivePrefix = {arXiv},
       eprint = {2207.01946},
 primaryClass = {astro-ph.IM},
       adsurl = {https://ui.adsabs.harvard.edu/abs/2023A&A...674A..22G}
}

@ARTICLE{McQuillan2014,
       author = {{McQuillan}, A. and {Mazeh}, T. and {Aigrain}, S.},
        title = "{Rotation Periods of 34,030 Kepler Main-sequence Stars: The Full Autocorrelation Sample}",
      journal = {\apjs},
         year = 2014,
        month = apr,
       volume = {211},
       number = {2},
          eid = {24},
        pages = {24},
          doi = {10.1088/0067-0049/211/2/24},
archivePrefix = {arXiv},
       eprint = {1402.5694},
 primaryClass = {astro-ph.SR},
       adsurl = {https://ui.adsabs.harvard.edu/abs/2014ApJS..211...24M}
}

@ARTICLE{2020NumPy-Array,
  author  = {Harris, Charles R. and Millman, K. Jarrod and
            van der Walt, Stéfan J and Gommers, Ralf and
            Virtanen, Pauli and Cournapeau, David and
            Wieser, Eric and Taylor, Julian and Berg, Sebastian and
            Smith, Nathaniel J. and Kern, Robert and Picus, Matti and
            Hoyer, Stephan and van Kerkwijk, Marten H. and
            Brett, Matthew and Haldane, Allan and
            Fernández del Río, Jaime and Wiebe, Mark and
            Peterson, Pearu and Gérard-Marchant, Pierre and
            Sheppard, Kevin and Reddy, Tyler and Weckesser, Warren and
            Abbasi, Hameer and Gohlke, Christoph and
            Oliphant, Travis E.},
  title   = {Array programming with {NumPy}},
  journal = {Nature},
  year    = {2020},
  volume  = {585},
  pages   = {357–362},
  doi     = {10.1038/s41586-020-2649-2}
}

@article{Milliman_2015,
   title={BARIUM SURFACE ABUNDANCES OF BLUE STRAGGLERS IN THE OPEN CLUSTER NGC 6819},
   volume={150},
   ISSN={1538-3881},
   url={http://dx.doi.org/10.1088/0004-6256/150/3/84},
   DOI={10.1088/0004-6256/150/3/84},
   number={3},
   journal={The Astronomical Journal},
   publisher={American Astronomical Society},
   author={Milliman, Katelyn E. and Mathieu, Robert D. and Schuler, Simon C.},
   year={2015},
   month=aug, pages={84} }

@article{Milliman_2014,
   title={WIYN OPEN CLUSTER STUDY. LX. SPECTROSCOPIC BINARY ORBITS IN NGC 6819},
   volume={148},
   ISSN={1538-3881},
   url={http://dx.doi.org/10.1088/0004-6256/148/2/38},
   DOI={10.1088/0004-6256/148/2/38},
   number={2},
   journal={The Astronomical Journal},
   publisher={American Astronomical Society},
   author={Milliman, Katelyn E. and Mathieu, Robert D. and Geller, Aaron M. and Gosnell, Natalie M. and Meibom, Søren and Platais, Imants},
   year={2014},
   month=jul, pages={38} }

@ARTICLE{Andrae2023,
       author = {{Andrae}, R. and {Fouesneau}, M. and {Sordo}, R. and {Bailer-Jones}, C.~A.~L. and {Dharmawardena}, T.~E. and {Rybizki}, J. and {De Angeli}, F. and {Lindstr{\o}m}, H.~E.~P. and {Marshall}, D.~J. and {Drimmel}, R. and {Korn}, A.~J. and {Soubiran}, C. and {Brouillet}, N. and {Casamiquela}, L. and {Rix}, H.-W. and {Abreu Aramburu}, A. and {{\'A}lvarez}, M.~A. and {Bakker}, J. and {Bellas-Velidis}, I. and {Bijaoui}, A. and {Brugaletta}, E. and {Burlacu}, A. and {Carballo}, R. and {Chaoul}, L. and {Chiavassa}, A. and {Contursi}, G. and {Cooper}, W.~J. and {Creevey}, O.~L. and {Dafonte}, C. and {Dapergolas}, A. and {de Laverny}, P. and {Delchambre}, L. and {Demouchy}, C. and {Edvardsson}, B. and {Fr{\'e}mat}, Y. and {Garabato}, D. and {Garc{\'\i}a-Lario}, P. and {Garc{\'\i}a-Torres}, M. and {Gavel}, A. and {Gomez}, A. and {Gonz{\'a}lez-Santamar{\'\i}a}, I. and {Hatzidimitriou}, D. and {Heiter}, U. and {Jean-Antoine Piccolo}, A. and {Kontizas}, M. and {Kordopatis}, G. and {Lanzafame}, A.~C. and {Lebreton}, Y. and {Licata}, E.~L. and {Livanou}, E. and {Lobel}, A. and {Lorca}, A. and {Magdaleno Romeo}, A. and {Manteiga}, M. and {Marocco}, F. and {Mary}, N. and {Nicolas}, C. and {Ordenovic}, C. and {Pailler}, F. and {Palicio}, P.~A. and {Pallas-Quintela}, L. and {Panem}, C. and {Pichon}, B. and {Poggio}, E. and {Recio-Blanco}, A. and {Riclet}, F. and {Robin}, C. and {Santove{\~n}a}, R. and {Sarro}, L.~M. and {Schultheis}, M.~S. and {Segol}, M. and {Silvelo}, A. and {Slezak}, I. and {Smart}, R.~L. and {S{\"u}veges}, M. and {Th{\'e}venin}, F. and {Torralba Elipe}, G. and {Ulla}, A. and {Utrilla}, E. and {Vallenari}, A. and {van Dillen}, E. and {Zhao}, H. and {Zorec}, J.},
        title = "{Gaia Data Release 3. Analysis of the Gaia BP/RP spectra using the General Stellar Parameterizer from Photometry}",
      journal = {\aap},
         year = 2023,
        month = jun,
       volume = {674},
          eid = {A27},
        pages = {A27},
          doi = {10.1051/0004-6361/202243462},
archivePrefix = {arXiv},
       eprint = {2206.06138},
 primaryClass = {astro-ph.SR},
       adsurl = {https://ui.adsabs.harvard.edu/abs/2023A&A...674A..27A}
}

@ARTICLE{Torres2010,
       author = {{Torres}, G. and {Andersen}, J. and {Gim{\'e}nez}, A.},
        title = "{Accurate masses and radii of normal stars: modern results and applications}",
      journal = {\aapr},
         year = 2010,
        month = feb,
       volume = {18},
       number = {1-2},
        pages = {67-126},
          doi = {10.1007/s00159-009-0025-1},
archivePrefix = {arXiv},
       eprint = {0908.2624},
 primaryClass = {astro-ph.SR},
       adsurl = {https://ui.adsabs.harvard.edu/abs/2010A&ARv..18...67T}
}

@ARTICLE{Chaplin2013,
       author = {{Chaplin}, William J. and {Miglio}, Andrea},
        title = "{Asteroseismology of Solar-Type and Red-Giant Stars}",
      journal = {\araa},
         year = 2013,
        month = aug,
       volume = {51},
       number = {1},
        pages = {353-392},
          doi = {10.1146/annurev-astro-082812-140938},
archivePrefix = {arXiv},
       eprint = {1303.1957},
 primaryClass = {astro-ph.SR},
       adsurl = {https://ui.adsabs.harvard.edu/abs/2013ARA&A..51..353C}
}

@article{Yang_2012,
   title={WIYN OPEN CLUSTER STUDY II: WIDE-FIELD CCD PHOTOMETRY OF THE OLD OPEN CLUSTER NGC 6819},
   volume={762},
   ISSN={1538-4357},
   url={http://dx.doi.org/10.1088/0004-637X/762/1/3},
   DOI={10.1088/0004-637x/762/1/3},
   number={1},
   journal={The Astrophysical Journal},
   publisher={American Astronomical Society},
   author={Yang, Soung-Chul and Sarajedini, Ata and Deliyannis, Constantine P. and Sarrazine, Angela R. and Kim, Sang Chul and Kyeong, Jaemann},
   year={2012},
   month=dec, pages={3} }

@article{Anthony_Twarog_2014,
   title={AuvbyCaHβ ANALYSIS OF THE OLD OPEN CLUSTER, NGC 6819},
   volume={148},
   ISSN={1538-3881},
   url={http://dx.doi.org/10.1088/0004-6256/148/3/51},
   DOI={10.1088/0004-6256/148/3/51},
   number={3},
   journal={The Astronomical Journal},
   publisher={American Astronomical Society},
   author={Anthony-Twarog, Barbara J. and Deliyannis, Constantine P. and Twarog, Bruce A.},
   year={2014},
   month=aug, pages={51} }

@article{Zwicker_2024,
   title={Investigating Mass Segregation of the Binary Stars in the Open Cluster NGC 6819},
   volume={967},
   ISSN={1538-4357},
   url={http://dx.doi.org/10.3847/1538-4357/ad39c6},
   DOI={10.3847/1538-4357/ad39c6},
   number={1},
   journal={The Astrophysical Journal},
   publisher={American Astronomical Society},
   author={Zwicker, Claire and Geller, Aaron M. and Childs, Anna C. and Motherway, Erin and von Hippel, Ted},
   year={2024},
   month=may, pages={44} }

@article{Kalirai_2001,
   title={The CFHT Open Star Cluster Survey. II. Deep CCD Photometry of the Old Open Star Cluster NGC 6819},
   volume={122},
   ISSN={0004-6256},
   url={http://dx.doi.org/10.1086/321141},
   DOI={10.1086/321141},
   number={1},
   journal={The Astronomical Journal},
   publisher={American Astronomical Society},
   author={Kalirai, Jasonjot Singh and Richer, Harvey B. and Fahlman, Gregory G. and Cuillandre, Jean-Charles and Ventura, Paolo and D’Antona, Francesca and Bertin, Emmanuel and Marconi, Gianni and Durrell, Patrick R.},
   year={2001},
   month=jul, pages={266–282} }

@MISC{keplerhandbook,
       author = {{Van Cleve}, Jeffrey E. and {Christiansen}, Jessie L. and {Jenkins}, Jon M. and {Caldwell}, Douglas A. and {Barclay}, Thomas and {Bryson}, Stephen T. and {Burke}, Christopher J. and {Cambell}, Jennifer and {Catanzarite}, Joseph and {Clarke}, Bruce D. and {Coughlin}, Jeffrey L. and {Girouard}, F. and {Haas}, Michael R. and {Klaus}, Todd C. and {Kolodziejczak}, Jeffrey J. and {Li}, Jie and {McCauliff}, Sean D. and {Morris}, Robert L. and {Mullally}, Fergal and {Quintana}, Elisa V. and {Rowe}, Jason and {Sabale}, Anima and {Seader}, Shawn and {Smith}, Jeffrey C. and {Still}, Martin D. and {Tenenbaum}, Peter G. and {Thompson}, Susan E. and {Twicken}, Joesph D. and {Kamal Uddin}, Akm and {Zamudio}, Khadeejah},
        title = "{Kepler Data Characteristics Handbook}",
 howpublished = {Kepler Science Document KSCI-19040-005, id. 2. Edited by Doug Caldwell, Jon M. Jenkins, Michael R. Haas and Natalie Batalha},
         year = 2016,
        month = dec,
          eid = {2},
        pages = {2},
       adsurl = {https://ui.adsabs.harvard.edu/abs/2016ksci.rept....2V}
}

@ARTICLE{Jenkins2010,
       author = {{Jenkins}, Jon M. and {Caldwell}, Douglas A. and {Chandrasekaran}, Hema and {Twicken}, Joseph D. and {Bryson}, Stephen T. and {Quintana}, Elisa V. and {Clarke}, Bruce D. and {Li}, Jie and {Allen}, Christopher and {Tenenbaum}, Peter and {Wu}, Hayley and {Klaus}, Todd C. and {Middour}, Christopher K. and {Cote}, Miles T. and {McCauliff}, Sean and {Girouard}, Forrest R. and {Gunter}, Jay P. and {Wohler}, Bill and {Sommers}, Jeneen and {Hall}, Jennifer R. and {Uddin}, AKM K. and {Wu}, Michael S. and {Bhavsar}, Paresh A. and {Van Cleve}, Jeffrey and {Pletcher}, David L. and {Dotson}, Jessie A. and {Haas}, Michael R. and {Gilliland}, Ronald L. and {Koch}, David G. and {Borucki}, William J.},
        title = "{Overview of the Kepler Science Processing Pipeline}",
      journal = {\apjl},
         year = 2010,
        month = apr,
       volume = {713},
       number = {2},
        pages = {L87-L91},
          doi = {10.1088/2041-8205/713/2/L87},
archivePrefix = {arXiv},
       eprint = {1001.0258},
 primaryClass = {astro-ph.EP},
       adsurl = {https://ui.adsabs.harvard.edu/abs/2010ApJ...713L..87J}
}

@ARTICLE{vizier,
       author = {{Ochsenbein}, F. and {Bauer}, P. and {Marcout}, J.},
        title = "{The VizieR database of astronomical catalogues}",
      journal = {\aaps},
         year = 2000,
        month = apr,
       volume = {143},
        pages = {23-32},
          doi = {10.1051/aas:2000169},
archivePrefix = {arXiv},
       eprint = {astro-ph/0002122},
 primaryClass = {astro-ph},
       adsurl = {https://ui.adsabs.harvard.edu/abs/2000A&AS..143...23O}
}

@article{pal,
author = {P\'al, Andr\'as},
title = {fitsh– a software package for image processing},
journal = {\mnras},
volume = {421},
number = {3},
pages = {1825-1837},
year = {2012},
doi = {10.1111/j.1365-2966.2011.19813.x},
eprint = {/oup/backfile/content_public/journal/mnras/421/3/10.1111/j.1365-2966.2011.19813.x/2/mnras0421-1825.pdf}
}

@InProceedings{Pandas,
  author    = { Wes McKinney },
  title     = { Data Structures for Statistical Computing in Python },
  booktitle = { Proceedings of the 9th Python in Science Conference },
  pages     = { 51 - 56 },
  year      = { 2010 },
  editor    = { St\'efan van der Walt and Jarrod Millman }
}

@ARTICLE{SoaresFurtado2017,
   author = {{Soares-Furtado}, M. and {Hartman}, J.~D. and {Bakos}, G.~{\'A}. and 
	{Huang}, C.~X. and {Penev}, K. and {Bhatti}, W.},
    title = "{Image Subtraction Reduction of Open Clusters M35 $\&$ NGC 2158 in the K2 Campaign 0 Super Stamps}",
  journal = {\pasp},
archivePrefix = "arXiv",
   eprint = {1703.00030},
 primaryClass = "astro-ph.IM",
     year = 2017,
    month = apr,
   volume = 129,
   number = 4,
    pages = {044501},
      doi = {10.1088/1538-3873/aa5c7c},
   adsurl = {http://adsabs.harvard.edu/abs/2017PASP..129d4501S}
}

@ARTICLE{vartools,
   author = {{Hartman}, J.~D. and {Bakos}, G.~{\'A}.},
    title = "{VARTOOLS: A program for analyzing astronomical time-series data}",
  journal = {Astronomy and Computing},
archivePrefix = "arXiv",
   eprint = {1605.06811},
 primaryClass = "astro-ph.IM",
     year = 2016,
    month = oct,
   volume = 17,
    pages = {1-72},
      doi = {10.1016/j.ascom.2016.05.006},
   adsurl = {http://adsabs.harvard.edu/abs/2016A%26C....17....1H}
}

@ARTICLE{Watson2006,
   author = {{Watson}, C.~L. and {Henden}, A.~A. and {Price}, A.},
    title = "{The International Variable Star Index (VSX)}",
  journal = {Society for Astronomical Sciences Annual Symposium},
     year = 2006,
    month = may,
   volume = 25,
    pages = {47},
   adsurl = {http://adsabs.harvard.edu/abs/2006SASS...25...47W}
}

@ARTICLE{Watson2017,
   author = {{Watson}, C. and {Henden}, A.~A. and {Price}, A.},
    title = "{VizieR Online Data Catalog: AAVSO International Variable Star Index VSX (Watson+, 2006-2014)}",
  journal = {VizieR Online Data Catalog},
     year = 2017,
    month = may,
   volume = 1,
   adsurl = {http://adsabs.harvard.edu/abs/2017yCat....102027W}
}

@ARTICLE{1967PZ.....16..191B,
       author = {{Barkhatova}, K.~A. and {Vasilevsky}, A.~E.},
        title = "{On a Possible Variable Star in an Open Stellar Cluster NGC 6819}",
      journal = {Peremennye Zvezdy},
         year = 1967,
        month = jan,
       volume = {16},
        pages = {191},
       adsurl = {https://ui.adsabs.harvard.edu/abs/1967PZ.....16..191B}
}

@MISC{lightkurve,
   author = {{Lightkurve Collaboration} and {Cardoso}, J.~V.~d.~M. and
             {Hedges}, C. and {Gully-Santiago}, M. and {Saunders}, N. and
             {Cody}, A.~M. and {Barclay}, T. and {Hall}, O. and
             {Sagear}, S. and {Turtelboom}, E. and {Zhang}, J. and
             {Tzanidakis}, A. and {Mighell}, K. and {Coughlin}, J. and
             {Bell}, K. and {Berta-Thompson}, Z. and {Williams}, P. and
             {Dotson}, J. and {Barentsen}, G.},
    title = "{Lightkurve: Kepler and TESS time series analysis in Python}",
howpublished = {Astrophysics Source Code Library},
     year = 2018,
    month = dec,
archivePrefix = "ascl",
   eprint = {1812.013},
   adsurl = {http://adsabs.harvard.edu/abs/2018ascl.soft12013L},
}

@ARTICLE{1991A&A...251...49M,
       author = {{Manteiga}, M. and {Martinez Roger}, C. and {Morales}, C. and {Sabau}, L.},
        title = "{Blue stragglers : a search for binaries in the infrared.}",
      journal = {\aap},
         year = 1991,
        month = nov,
       volume = {251},
        pages = {49},
       adsurl = {https://ui.adsabs.harvard.edu/abs/1991A&A...251...49M}
}

@ARTICLE{1988AJ.....95..785K,
       author = {{Kaluzny}, Janusz and {Shara}, Michael M.},
        title = "{A CCD Survey for Contact Binaries in Six Open Clusters}",
      journal = {\aj},
         year = 1988,
        month = mar,
       volume = {95},
        pages = {785},
          doi = {10.1086/114677},
       adsurl = {https://ui.adsabs.harvard.edu/abs/1988AJ.....95..785K}
}

@ARTICLE{Sanjayan2022,
       author = {{Sanjayan}, S. and {Baran}, A.~S. and {N{\'e}meth}, P. and {Kinemuchi}, K.},
        title = "{Variable Star Population in the Open Cluster NGC 6819 Observed by the Kepler Spacecraft}",
      journal = {\actaa},
         year = 2022,
        month = jan,
       volume = {72},
       number = {4},
        pages = {267-295},
          doi = {10.32023/0001-5237/72.4.3},
archivePrefix = {arXiv},
       eprint = {2405.03887},
 primaryClass = {astro-ph.SR},
       adsurl = {https://ui.adsabs.harvard.edu/abs/2022AcA....72..267S}
}

@ARTICLE{Kristiansen2022,
       author = {{Kristiansen}, Martti H.~K. and {Rappaport}, Saul A. and {Vanderburg}, Andrew M. and {Jacobs}, Thomas L. and {Schwengeler}, Hans Martin and {Gagliano}, Robert and {Terentev}, Ivan A. and {LaCourse}, Daryll M. and {Omohundro}, Mark R. and {Schmitt}, Allan R. and {Powell}, Brian P. and {Kostov}, Veselin B.},
        title = "{The Visual Survey Group: A Decade of Hunting Exoplanets and Unusual Stellar Events with Space-Based Telescopes}",
      journal = {arXiv e-prints},
         year = 2022,
        month = may,
          eid = {arXiv:2205.07832},
        pages = {arXiv:2205.07832},
archivePrefix = {arXiv},
       eprint = {2205.07832},
 primaryClass = {astro-ph.EP},
       adsurl = {https://ui.adsabs.harvard.edu/abs/2022arXiv220507832K}
}

@ARTICLE{2002MNRAS.330..737S,
       author = {{Street}, R.~A. and {Horne}, Keith and {Lister}, T.~A. and {Penny}, A. and {Tsapras}, Y. and {Quirrenbach}, A. and {Safizadeh}, N. and {Cooke}, J. and {Mitchell}, D. and {Collier Cameron}, A.},
        title = "{Variable stars in the field of open cluster NGC 6819}",
      journal = {\mnras},
         year = 2002,
        month = mar,
       volume = {330},
       number = {3},
        pages = {737-754},
          doi = {10.1046/j.1365-8711.2002.05125.x},
       adsurl = {https://ui.adsabs.harvard.edu/abs/2002MNRAS.330..737S}
}

@ARTICLE{1971IBVS..606....1L,
       author = {{Lindoff}, U.},
        title = "{A Late Type Variable in NGC 6819}",
      journal = {Information Bulletin on Variable Stars},
         year = 1971,
        month = dec,
       volume = {606},
        pages = {1},
       adsurl = {https://ui.adsabs.harvard.edu/abs/1971IBVS..606....1L}
}
\bibliographystyle{aasjournalv7}

\end{document}